\documentclass[%
 reprint,
 amsmath,amssymb,
 aps,
]{revtex4-2}

\usepackage{dcolumn}
\usepackage{bm}

\usepackage{graphicx}
\usepackage{subfigure}
\usepackage{graphicx,hyperref}
\usepackage[normalem]{ulem}
\expandafter\let\csname equation*\endcsname\relax
\expandafter\let\csname endequation*\endcsname\relax
\usepackage{amsmath}
\usepackage{amssymb}
\usepackage{multirow, makecell, comment} 
\newcommand{\fla}[1]{\begin{flalign}#1\end{flalign}}
\usepackage{braket}
\usepackage{lineno}
\usepackage{lipsum}
\usepackage{xcolor}
\usepackage{soul}
\usepackage{array} 

\newcommand{\tmYAG}{Tm$^{3+}$:Y$_3$Al$_5$O$_{12}$\,}
\newcommand{\YAG}{Y$_3$Al$_5$O$_{12}$\,}
\newcommand{\tmtrans}{$^3$H$_6 (0)\rightarrow ^3$H$_4 (0)$\,}

\begin{document}

\preprint{APS/123-QED}

\title{Pulse area theorem in a single mode waveguide and its application to photon echo and optical memory in \tmYAG}

\author{  S.A. Moiseev$^{1}$}
\email{s.a.moiseev@kazanqc.org}
\author{M.M. Minnegaliev$^{1}$,
E.S. Moiseev $^{1}$,
K.I. Gerasimov$^{1}$,
A.V. Pavlov$^{1}$,   T.A. Rupasov$^{1}$, N.N. Skryabin$^{2}$, A.A. Kalinkin$^{2}$, S.P. Kulik$^{2,3}$}

\affiliation{$^{1}$Kazan Quantum Center, Kazan National Research Technical University n.a. A.N. Tupolev-KAI, 10 K. Marx St., 420111, Kazan, Russia}
\affiliation{$^{2}$ Quantum Technologies Center and Faculty of Physics, M.V. Lomonosov Moscow State University, Leninskie Gory, 119991, Moscow, Russia}
\affiliation{$^{3}$ Laboratory of quantum engineering of light, South Ural State University (SUSU),	 Lenin Avenue, 454080, Chelyabinsk, Russia}

\date{\today}

\begin{abstract}
We derive the area theorem for light pulses interacting with inhomogeneously broadened ensemble of two-level atoms in a single-mode optical waveguide and present its analytical solution for Gaussian-type modes, which demonstrates the significant difference from the formation of $2\pi$ pulses by plane waves.
We generalize this theorem to the description of photon echo and apply it to the two-pulse (primary) echo and the revival of silenced echo (ROSE) protocol of photon echo quantum memory.
For the first time, we implemented ROSE protocol in a single-mode laser-written waveguide made of an optically thin crystal \tmYAG.
The experimental data obtained are satisfactorily explained by the developed theory.
Finally, we discuss the obtained experimental results  and possible applications of the derived pulse area approach.
\end{abstract}

                              
\maketitle

\section{\label{sec:level1}Introduction}
The coherent interaction of a light pulse with resonant atomic ensembles plays a significant role in modern optics and quantum technologies \cite{allen1975,Scully1997, BoydNonLinOptics,Sangouard:11}.
These interactions often poses nonlinear character, study of which is a difficult theoretical task.
The pulse area theorem \cite{McCall1969} provides a simple but powerful tool for general analysis of nonlinear coherent light-atoms dynamics in  self-induced transparency \cite{McCall1969}, optical solitons \cite{Eberly2016}, superradiance \cite{Greiner-2001}, photon echo in optically dense media  \cite{HAHN1971, Moiseev2020, Moiseev2022}, to name a few.
The approach was developed for propagating plane light waves that interact with atoms in free space.
Recent progress in integrated quantum photonics \cite{Kim:20,Hibino-2003,Rodenas-2011,Okhrimchuk-2005} 
motivates the study of the coherent interaction between light pulses and resonant atomic ensembles in optical waveguides, where the development of waveguide optical quantum memory (QM) attracts growing attention \cite{Sinclair-2010, Saglamyurek2011, marzban-PRL-2015,Corrielli-PRApplied-2016, Liu-2020,Wang_2022}.

The goal of an optical QM is to store quantum states of light for subsequent on-demand retrieval at an arbitrary time \cite{QMReview, Hammerer2010,Tittel2010,Bussieres:13,Khabat2016,CHANELIERE2018}. QM is a vital component for numerous quantum technologies, such as long-distance quantum communications \cite{Sangouard:11,S-Wehner-2018,Wang-2020}, quantum state preparation \cite{Kaneda:17}, and a synchronization unit for optical quantum processing \cite{Bussieres:13}.

Great promises are associated with the  photon-echo-based  optical QM in crystals doped with \textit{rare earth ions} (REI) \cite{Tittel2010} that have a long lifetime of quantum coherence at optical and microwave transitions.
Such optical QM has advantages in its multiplexing capacity for storing a large number of temporary light modes and demonstrates high efficiency in REIs-doped crystals, for example, 58\% in the cavity assistant scheme \cite{2013-Sabooni-PRL} and 69\% in the free space \cite {2010-Nature-Hedges}, which are comparable to the 76\% efficiency of single mode storage achieved with QM protocol based on electromagnetically induced transparency in REI-doped crystal \cite{Schraft_2016}.

Currently, there is growing interest in the implementation of photon echo QM in optical waveguides \cite{Saglamyurek2011,Thiel_2012,Saglamyurek2015,Corrielli-PRApplied-2016, Optica-Liu:20} doped by REIs 
which seem as a convenient platform for implementation of on-chip QM.
Quantum storage in REI-doped waveguides was demonstrated in experiments on heralded single photon storage \cite{Seri:18,Askarani2019}, on-demand qubit storage \cite{Liu-2020} and frequency-multiplexed  storage \cite{Seri:19}.
All of the above  experiments are based on the scheme of reversible photon echo in an optically dense medium \cite{Moiseev2001,Moiseev2003,Kraus2006}, realized for inhomogeneous broadening in the form of a periodic narrow \textit{atomic frequency  combs} that is called AFC protocol \cite{Afzelius:09}.
There is a particular interest in the \textit{revival of silenced echo} (ROSE) protocol \cite{Damon2011, 2021-PRB-Minnegaliev} for implementation of 
optically controlled on-demand QM   with low quantum noise background in atomic ensemble with naturally inhomogeneous broadening and narrow homogeneous lines that is embedded in an optical waveguide.

Recently  the ROSE protocol  was  implemented in the $^{151}\text{Eu}^{3+}:\text{Y}_2 \text{Si} \text{O}_5$ crystal with the type-II single mode laser-written waveguide \cite{Optica-Liu:20}. 
This type of the waveguide supports propagation of light modes with only one polarization.
At the same time, light modes of arbitrary polarization can propagate in a type-III waveguide.
Such type-III depressed cladding single-mode waveguides were fabricated by femtosecond laser writing technique in the crystal \tmYAG \cite{Skryabin20}. 
Therefore, it is desirable to implement the ROSE protocol in this waveguide, which is the subject of experimental studies of this work.

The mentioned need for a highly efficient implementation of photon echo QMs and its integration with waveguide schemes makes it relevant to develop a general theoretical approach for describing the coherent interaction between light and atoms in optical waveguides. 
The approach should properly simulate the application of intense control laser fields, e.g., $\pi-$pulses, that induce nonlinear dynamics in resonant atoms.

Since the McCall-Hahn work \cite{McCall1969}, it was discovered that the most general patterns of linear and nonlinear interaction between light pulses and coherent two-level media can be described by the area theorem
\cite{allen1975,Greiner-2001, Eberly2016}.
The area theorem was also applied to the description of photon echo in optically dense media \cite{HAHN1971, Moiseev1987, Urmancheev2019,Moiseev2020}, in Fabry-Perot resonator \cite{Moiseev2022} as well as it was applied for studies of photon echo CRIB  protocol in free space \cite{Moiseev2004b} and for description of cavity assistant ROSE protocol \cite{2021-PRB-Minnegaliev}. 
It is worth noting that in recent experiments \cite{Urmancheev2019,2021-PRB-Minnegaliev}, it has been found that in media with a symmetrical form of inhomogeneous broadening, the behavior of the pulse area of the echo signal almost flatly reproduces the behavior of the amplitude of echo signals. 
It makes this approach universal for describing linear and nonlinear behaviour in various scenario of photon echo, as well as photon echo based QM protocols in optically dense resonant media.

In this work, we derive the \textit{waveguide pulse area} (WPA) theorem  for the resonant interaction between light pulse and two-level atoms in a single-mode optical waveguide 
(see Fig. \ref{Figure_1}).
We found the analytical form of the WPA theorem, demonstrating significant differences in a formation of stable $2\pi$ pulses compared to the well-known McCall-Hahn pulse area theorem \cite{McCall1969}.
Then we generalize WPA theorem for the echo signal emission and apply it for description of the two-pulse (primary) photon echo and ROSE protocol of photon echo QM.
For the first time, we  present experimental results on ROSE protocol in a laser-written waveguide of a Tm$^{3+}$ crystal:Y$_3$Al$_5$O$_{12}$.
Then we discuss the obtained experimental data using the WPA theorem, focusing on the factors leading to a negative impact on the efficiency of the QM protocol implementation and outlining the possible applications of the developed theoretical approach to other problems.

\section{Waveguide pulse area theorem}
\label{sec::math_formalism}
\subsection{Maxwell-Bloch equations in single mode waveguide and pulse area theorem}

We derive the equations of motion by starting with the Hamiltonian $\hat{H}_0$, which is composed of
Hamiltonian of non-interacting two-level atoms $\hat{H}_a$ with inhomogeneously broadened resonant transition, Hamiltonian of waveguide light modes $\hat{H}_f$, and an interaction Hamiltonian  between the atoms and waveguide modes in dipole and rotating wave approximations $\hat{V}_{fa}$:
\fla{
\hat{H}_0=\hat{H}_a+\hat{H}_f+\hat{V}_{fa}. 
\label{Hamiltonian}
}
We consider the free atomic Hamiltonian to contain several types of two-level atoms with different dipole moments of resonant transition $\mathbf{d}_m$.
It is a typical case with rare-earth ions, where active ions may substitute for host atoms at different positions in a crystal, leading to several ($M$) groups of atoms with different dipole moments \cite{Goldner_2015}.
The free atomic Hamiltonian is
\fla{
\hat{H}_a=\sum_{m=1}^{M}\sum_{j=1}^{N_m} \frac{\hbar}{2} (\omega_0+\Delta_j)\sigma_{3;m} ^j, 
}
where 
$N_m$ is a number of atoms within m-th group, $\omega_0$ is the carrier frequency of the light field coinciding with the center of the line of the optical atomic transition (see  Fig. \ref{Figure_1}), $\sigma^{j}_{1,m},\sigma^{j}_{2,m},\sigma^{j}_{3,m}$ are standard set of Pauli matrices related to the 
$j$-th two-level atom from m-th group, $\Delta_j$ is detuning of j-th atom from the carrier frequency.

The Hamiltonian of the waveguide light mode propagating along the $z$ direction is \cite{Shen_2009,Moiseev_2010,Roy2017}
\fla{
\hat{H}_f=\hbar  \int dz a^{\dagger}(z)[\omega_0 a(z) -i v_g \frac{\partial}{\partial z}a(z)],
} 
where $v_g= \frac{\partial \omega_0}{\partial \beta}$ is a group velocity (see  Fig. \ref{Figure_1}), 
 $\beta=\sqrt{k^2(\omega_0)-(2\pi/\Lambda)^2}$, 
 $k(\omega)=n(\omega)\omega/c$, $n(\omega)=\sqrt{\varepsilon (\omega)}$,
$\varepsilon(\omega)$ is an electric permittivity of  dielectric medium
and $\Lambda$ being a critical waveguide wavelength \cite{Kogelnik1988},
$a(z)$ ($a^{\dagger}(z)$) is bosonic annihilation (creation) operator of light field mode  at given coordinate $z$  with commutation relation $\left[ \hat{a}(z), \hat{a}^{\dagger}(z') \right] = \delta(z-z')$. 
The operators are expressed with a help of their one dimensional Fourier image $a(k)$:
\fla{
\hat{a}(z) =\frac{1}{\sqrt{2\pi}}\int d k e^{i(k-\beta) z} a(k),
}
The electric field of the waveguide mode \cite{Huttner_1992}, taking into account their structure in an optical waveguide, has the form:

\begin{figure*}
\centering
\includegraphics[width=0.8\linewidth]{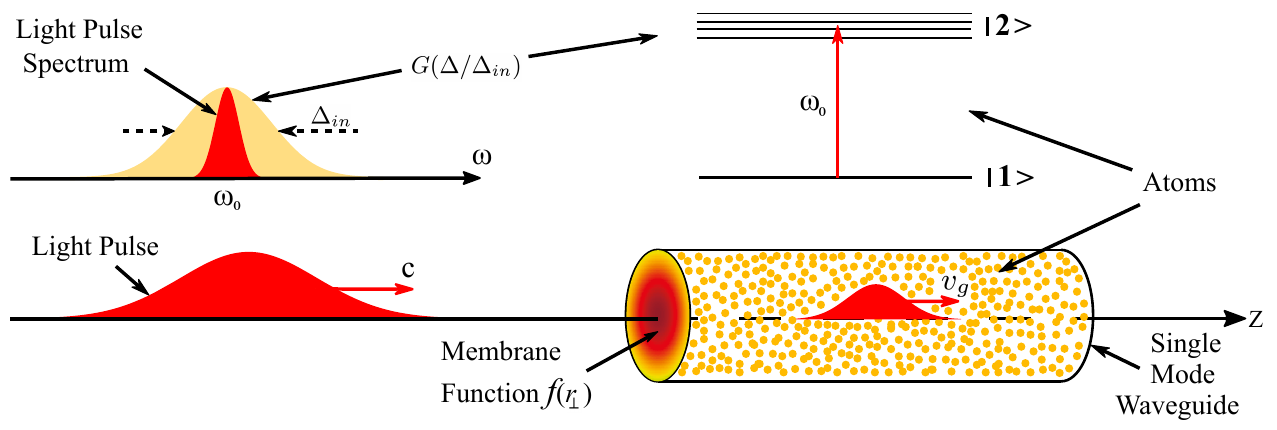}
\caption{Spatial scheme of interaction of light pulses with an ensemble of two-level atoms in a single-mode optical waveguide. $G(\Delta/\Delta_{in})$ is an inhomogeneous broadening with linewidth $\Delta_{in}$, $c$ and $v_g$ are the speeds of light in free space and in the waveguide; $f(\mathbf{r}_{\perp})$ is a membrane function  of light field in the the single mode waveguide.}
\label{Figure_1}
\end{figure*}

\fla{
\mathbf{\hat{E}}(\mathbf{r}) =  i\mathbf{e} E_0 f(\mathbf{r}_{\perp} ) \hat{a}(z)e^{i\beta z_j}   + h.c., 
\label{eq::Eoper}
}
\\
where 
$E_0f(\mathbf{r}_{\perp} )$ is a single photon electric field amplitude \cite{Scully1997} at the point $\mathbf{r}_{\perp}$ in the transverse plane of the single mode waveguide with reference frame being chosen as
$\mathbf{r}^{j}=\mathbf{r}^{j}_{\perp} + z_{j}\mathbf{e}_z$.
$f(\mathbf{r}_{\perp})$ is  a membrane function  of the light mode \cite{Kogelnik1988,okamoto2021fundamentals},
$\mathbf{e}$ is a polarization vector of the light field,
$E_0\cong(\frac{\hbar\omega_0}{2\varepsilon_0 \varepsilon (\omega_0) S})^{1/2}$, 
$\hbar$ is reduced Planck’s constant, $S$ is the cross-section of the light beam, $\varepsilon_0$ is the electric permittivity of vacuum,  respectively \cite{Scully1997}.
We assume that  permittivity of the dielectric medium is constant and does not change over the bandwidth of interest.

The  interaction Hamiltonian between the atoms and the waveguide mode in the rotating wave approximation takes form
\fla{
\hat{V}_{fa}=- \frac{\hbar}{2}\sum_{m=1}^{M} \sum_{j=1}^{N_{m}} \Omega_{0,m}(\mathbf{r}^{j}_{\perp}) \sigma^{j}_{-,m} \hat{a}^{\dagger}(z_j)e^{- i\beta z_j} +h.c.,
}
where $\sigma_{\pm;m}^j=\frac{1}{2}(\sigma_{1;m}^j\pm i\sigma_{2;m})$ are the isospin flip operators for j-th atom of m-th group, $\Omega_{0,m}(\mathbf{r}_{\perp}^j )=\Omega_{m} f(\mathbf{r}_{\perp}^j)$ is a  coupling constant between $j$-th atom of m-th group and waveguide mode with 
$\Omega_{m}=E_0 \langle \mathbf{d}_m\cdot \mathbf{e} \rangle/\hbar$
being a single photon Rabi-frequency of  $j$-th atom located at $\mathbf{r}^{j}$. 
The membrane function $f(\mathbf{r}_{\perp})$ determines the dependence of the interaction constant of an atom on its position $\mathbf{r_{\perp}}$ in the transverse plane of the waveguide.
For sake of simplicity but without losing generality, we assume $f(\mathbf{r}_{\perp})$ to be a real-valued function.

We introduce slowly varied operators 
\fla{
\hat{a}_p(z,t)&=i e^{i\omega_0 t - i\beta z} \hat{a}(z,t), \\
\hat{\tilde{\sigma}}^{j}_{-,m}(t)&= e^{i\omega_0 t  - i\beta z}\hat{\sigma}^{j}_{-,m}(t),
\label{eq::field}
\\
\hat{\tilde{\sigma}}^{j}_{+,m}(t)&= e^{-i\omega_0 t  + i\beta z}\hat{\sigma}^{j}_{+,m}(t),
}
with the same commutation relations as $\hat{a}(z,t)$,$\hat{\sigma}^{j}_{-,m}(t)$ and $\hat{\tilde{\sigma}}^{j}_{+,m}(t)$, respectively. 
The index $p$ in $a_{p}(z,t)$ indicates name of the studied pulse for further echo analysis, i.e.  $p=(s,1,2,e)$  correspond to signal, first control, second control and echo pulses, respectively.

The resulted Heisenberg equations for the slowly varied operators are
\fla{
&\left( \frac{\partial}{\partial t} + v_g \frac{\partial}{\partial z} \right) \hat{a}_p(z,t) =  
\nonumber\\ 
&\frac{i}{2}\sum_{m=1}^{M} \sum^{N_m}_{j=1} \Omega_{0,m} (\mathbf{r}^{j}_{\perp} ) \hat{\tilde{\sigma}}^{j}_{-,m}(t)\delta(z-z_j),
\label{eq::Asign}
\\
& \frac{\partial \hat{\tilde{\sigma}}^{j}_{-,m} }{\partial t} = - i\Delta_{j}  \hat{\tilde{\sigma}}^{j}_{-,m} 
-\frac{i}{2} \Omega_{0,m}(\mathbf{r}^{j}_{\perp})\hat{a}_p(z_j,t) \sigma^{j}_{3,m},
\label{eq::Asigma_-}
\\
&\frac{\partial \hat{\tilde{\sigma}}^{j}_{+,m} }{\partial t} =i\Delta_{j}  \hat{\tilde{\sigma}}^{j}_{+,m} 
+\frac{i}{2}\Omega_{0,m}(\mathbf{r}^{j}_{\perp})\hat{a}_p^{\dagger}(z_j,t)\sigma^{j}_{3,m},
\label{eq::Asigma_+}
 \\
&\frac{\partial \sigma^{j}_{3,m} }{\partial t} =  - i\Omega_{0,m}(\mathbf{r}^{j}_{\perp})\left[\hat{a}_p^{\dagger}(z_j,t) \hat{\tilde{\sigma}}^{j}_{-,m} -\hat{a}_p(z_j,t)\hat{\tilde{\sigma}}^{j}_{+,m}\right].
\label{eq::Asigma_Z}
}

\noindent
It is worth noting, that at relatively large number of atoms with negligible small  change of the population ($\sigma^{j}_{3,m}(t)\cong \sigma^{j}_{3,m}(t_0)$), the system of  Eqs. \eqref{eq::Asign} - \eqref{eq::Asigma_Z} is linearized. 
Such conditions are often used in different photon echo quantum memory schemes \cite{Tittel2010, Moiseev_2013, Moiseev_2020}, where their solutions can be found for the arbitrary quantum states of lights.

The operator Eqs.(\ref{eq::Asign})-(\ref{eq::Asigma_Z}) can be simplified in classical limit of light field into c-number equations on average values of the operators by a proper splitting of two-particle correlators.
For an atomic ensemble with large $N_m$ and the waveguide mode being in coherent state  we replace the product of operator by a product of their mutual average values \cite{Haake_1979,allen1975}
\fla{
\langle \hat{a}^{(\dagger)}_p(z_j,t) \langle\sigma^{j}_{3,m}(t)\rangle \cong \langle \hat{a}^{(\dagger)}_p(z_j,t)\rangle \langle\sigma^{j}_{3,m}(t)\rangle, \\
\langle \hat{a}^{(\dagger)}_p(z_j,t) \sigma^{j}_{+(-),m}(t)\rangle \cong \langle \hat{a}^{(\dagger)}_p(z_j,t)\rangle \langle\sigma^{j}_{+(-),m}(t)\rangle.
}
For further convenience we switch to real-value equations by separating the constant phase shift $\phi_0$ and amplitude from the averaged field operator $\langle \hat{a}_0(z,t)\rangle=a_0(z,t)e^{i\phi_0}$ and including the phase shift in the atomic operators $\sigma^{j}_{+,m}(t)=e^{-i\phi_0}\sigma^{j}_{0;+,m}(t)$, 
$\sigma^{j}_{-,m}(t)=e^{i\phi_0}\sigma^{j}_{0;-,m}(t)$.
Finally, by switching to the components of Bloch vector for $j_m$ atom 
\fla{
v^{j}_{m}(t) &= i\langle(\sigma^{j}_{0;-,m}(t)-\sigma^{j}_{0;+,m}(t))\rangle, \\
u^{j}_{m}(t) &=\langle(\sigma^{j}_{0;-,m}(t)+\sigma^{j}_{0;+,m}(t))\rangle, \\
w^{j}_{m}(t) &=\langle \sigma^{j}_{3,m}(t)\rangle,
}
we get the following system of equations


 


\fla{
&\left( \frac{\partial}{\partial t} +\frac{\gamma_w}{2}+ v_g \frac{\partial}{\partial z} \right) a_{p}(z,t) =
\nonumber
\\
&
\sum_{m=1}^{M} \sum^{N_m}_{j=1} \frac{\Omega_{0,m}(\mathbf{r}^{j}_{\perp} )}{4}  v^{j}_{m}(t)\delta(z-z_j),
\label{eq::Asclass}
\\
&\frac{\partial u^{j}_{m}(t) }{\partial t} = -\frac{\gamma}{2}u^{j}_{m}(t) -  \Delta_{j}v^{j}_{m}(t),
\label{eq::u}
\\
&\frac{\partial v^{j}_{m}(t) }{\partial t} = -\frac{\gamma}{2}v^{j}_{m}(t) +
\Delta_{j}u^{j}_{m}(t)  +
\nonumber
\\
&\Omega_{0,m}(\mathbf{r}^{j}_{\perp}) a_{p}(z_j,t) w^{j}_{m}(t),
\label{eq::v}
 \\
&\frac{\partial w^{j}_{m}(t) }{\partial t} =  - \Omega_{0,m}(\mathbf{r}^{j}_{\perp})a_{p}(z_j,t) v^{j}_{m},
\label{eq::w}
}

\noindent
where we have added phenomenologically the decay constant $\gamma_w $ describing non-resonant losses of the waveguide modes and the decay constant of atomic phase relaxation
$\gamma=2/T_2$ (see Appendix A, where the decay constants $\gamma_w$ and $\gamma$ are introduced together with the related Langevin forces).
Although Eqs. \eqref{eq::Asclass}-\eqref{eq::w} do not  describe all the quantum properties of light and atoms, they are sufficient for analysing the efficiency, coherence, and spectral properties of optical QM for weak light fields.

Next, we derive a pulse area theorem for general description of the nonlinear properties of coherent interaction of light pulse with two-level medium in a single mode optical waveguide.
The presence of several dipole moments makes it difficult to define the total pulse area of the atomic ensemble. 
Hence instead of pulse area, we define \textit{envelope area} of the field that is agnostic to the presence of several dipole moments
\fla{
\theta_{p}(z)=\int_{t_o}^{t\gg \delta t_f} dt a_{p}(z,t).
}
We assume that pulse duration of the light pulses $\delta t_{f}$ is significantly shorter than the phase relaxation time of the atomic transition ($\gamma\delta t_f\ll1$).
By integrating  Eq. \eqref{eq::Asclass} over a  duration of the light pulse from its beginning $t_0$ to its end similarly to
\cite{allen1975,McCall1969} and using formal solution 
$r_m^j(t)=v^{j}_{m}(t)+iu^{j}_{m}(t)$ of Eqs. \eqref{eq::u}-\eqref{eq::w} with
 initial condition $v_m^j(t_0)=0, w_m^j(t_0)=w_0$,  
we get equation for the envelope area
\fla{
&\big(\frac{\partial}{\partial z}+\frac{\gamma_w}{2v_g}\big)\theta_{p}(z) =  
\nonumber
\\
&Re\{\sum_{m=1}^{M} \sum^{N_m}_{j=1} \frac{ \Omega_{0,m} (\mathbf{r}
^{j}_{\perp} )}{4v_g} \int_{t_o}^{t\gg \delta t_f}dtr^{j}_{m}(t)\}\delta(z-z_j)=
\nonumber
\\
& Re\{\sum_{m=1}^{M} \sum^{N_m}_{j=1} \frac{ \Omega_{0,m}^2 (\mathbf{r}
^{j}_{\perp} )}{4v_g}  \int_{t_o}^{t\gg \delta t_f}dt
\int_{t_o}^{t}dt'\cdot
\nonumber
\\
& e^{-(\gamma/2+i\Delta_j)(t-t')}
a_{p}(z_j,t') w^{j}_{m}(t')\}\delta(z-z_j).
\label{theta::1}
}

Taking into account that $a_{p}(z_j,t\gg \delta t_f)=0$, we can extend the limits of integration over time from  $\int_{t_0}^{t\gg \delta t_f}dt...$ to $\int_{t_0}^{\infty}dt...$
By changing the order of integration further
$\int_{t_0}^{\infty}dt\int_{t_0}^{t}dt'...\rightarrow $
$\int_{t_0}^{\infty}dt'\int_{t'}^{\infty}dt...$ and performing integration over $t$ we find

\fla{
&\big(\frac{\partial}{\partial z}+\frac{\gamma_w}{2v_g}\big)\theta_{p}(z) =  
 Re\{\sum_{m=1}^{M} \sum^{N_m}_{j=1} \frac{ \Omega_{0,m}^2 (\mathbf{r}
^{j}_{\perp} )}{4v_g} 
\frac{1}{\gamma/2+i\Delta_j}\cdot
\nonumber
\\
&\int_{t_o}^{\infty}dt' a_{p}(z_j,t') w^{j}_{m}(t')\}
\delta(z-z_j) =
\nonumber
\\
& \sum_{m=1}^{M} 
\sum^{N_m}_{j=1}  
\frac{\frac{1}{2}\gamma \Omega_{0,m} ^2(\mathbf{r}^{j}_{\perp} )}{4 v_g(\frac{1}{4}\gamma^2+\Delta_j^2)}
\int_{t_o}^{\infty}dt'
\alpha_{p}(z^j_m,t')w_m^j(t')\delta(z-z_j).
\label{eq::Asclass1}
}

\noindent

The coupling constant between an atom and light mode $\Omega_{0,m} (\mathbf{r}_{\perp} )$ is contained in the equations for the components of the Bloch vector \eqref{eq::Asclass}, \eqref{eq::v}, \eqref{eq::w}  
and in the equation for the envelope area \eqref{eq::Asclass1}. For further convinience the summation of an arbitrary function $F_m(\mathbf{r}^{j}_{\perp},z_j,\Delta_j,t)$ over the atoms in continuous limit may be approximated as an integration
\fla{
&\sum^{N_m}_{j=1} F_m(\mathbf{r}^{j}_{\perp}, z_j, \Delta_j,t) \delta(z-z_j)=
\nonumber
\\
&\rho_m\int d \Delta G(\frac{\Delta}{\Delta_{in}})\int_S dxdyF_m(\mathbf{r}_{\perp},z, \Delta,t),
\label{sum_int}
}

\noindent
where $S$ is a cross-section of the waveguide, 
$\rho_m=\frac{N_m}{LS}$ is a density of m-th atomic group with L being length the waveguide. For sake of simplicity we assume the identical inhomogeneous broadening profiles $G(\frac{\Delta}{\Delta_{in}})$ for different atomic groups with linewidth $\Delta_{in}$.
Substituting  Eq. \eqref{sum_int} in Eq. \eqref{eq::Asclass1} we get

\fla{
&\big(\frac{\partial}{\partial z}+\frac{\gamma_w}{2v_g}\big)\theta_{p}(z) =  
\nonumber
\\
& \sum_{m=1}^{M} 
\frac{\pi \rho_m}{4v_g} 
\int_S dx dy 
\Omega_{0,m}(\mathbf{r}_{\perp})
\int d \Delta
\frac{\frac{1}{2}\gamma  G(\frac{\Delta}{\Delta_{in}}) }{\pi(\frac{1}{4}\gamma^2+\Delta^2)}
\cdot
\nonumber
\\
&\Omega_{0,m} (\mathbf{r}_{\perp})\int_{t_o}^{\infty}dt'\alpha_{p}(z,t')
w_m(\mathbf{r}_{\perp},z,\Delta,t').
\label{theta:eq}
}
\\
For large enough inhomogeneous broadening ($\Delta_{in}\gg \gamma$) single atom spectral response is assumed to be delta function $\frac{\gamma}{\pi(\gamma^2+\Delta^2)}\cong\delta(\Delta)$.  If there is no any initial coherence  $v_m^j(t_0)=0$, the solution for resonant atomic coherence in Eqs. (\ref{eq::v}-\ref{eq::w}) is

\fla{
\Omega_{0,m} (\mathbf{r}_{\perp} )\int_{t_o}^{t\gg \delta t_f}dt' \alpha_{p}(z^j_m,t')
w_m(\mathbf{r}_{\perp},z,\Delta=0,t')
\nonumber \\
=v_m(\mathbf{r}_{\perp},z,\Delta=0,t)\vert_{t_0}^{t}=w_m^j(t_0)\sin\big( \Theta_{m,p}(\mathbf{r}_{\perp},z)\big),
} 
\noindent
where $\Theta_{m,p}(\mathbf{r}_{\perp},z)=
\Omega_{0,m} (\mathbf{r}_{\perp} )\theta_{p}(z)$ is a conventional pulse area of m-th atom located at a point with coordinates ($\mathbf{r}_{\perp},z$).
Substituting the initial condition for population $ w_m^j(t_0)=-1$ and using $\delta(\Delta)$
in  Eq. \eqref{theta:eq},  we get the following equation for the envelope area $\theta_{p}(z)$

\fla{
\big(\frac{\partial}{\partial z}+\frac{\gamma_w}{2v_g}\big)\theta_{p}(z)=-\Phi(\theta_{p}),
\label{Envelope_area}
}
where
\fla{
\Phi(\theta_{p})=\sum_{m=1}^{M}\frac{\xi_m}{2}\int_S dxdy  \Omega_{0,m} (\mathbf{r}_{\perp} )\sin\big( \Theta_{m,p}(\mathbf{r}_{\perp},z)\big).
\label{F_function}
}
\noindent
The atomic function
$\Phi(\theta_{p})$  describes the contribution of the total polarization within the fiber cross-section, 
$\xi_m=\frac{\pi  G(0) \rho_m}{2v_g}$, where
for large  $\Delta_{in}$ we can use $G(\frac{\Delta}{\Delta_{in}})=\frac{\Delta_{in}}{\pi(\Delta_{in}^2+\Delta^2)}$. 
The equation \eqref{Envelope_area} represents waveguide pulse area (WPA) theorem, which is an analog of the pulse area theorem for two-level atoms interacting in general waveguide mode. Below, we discuss several examples of the application of WPA theorem to Gaussian and Gaussian-type profiles of the waveguide modes.

\subsection{Gaussian mode}

\begin{figure*}
\centering
\includegraphics[width=\linewidth]{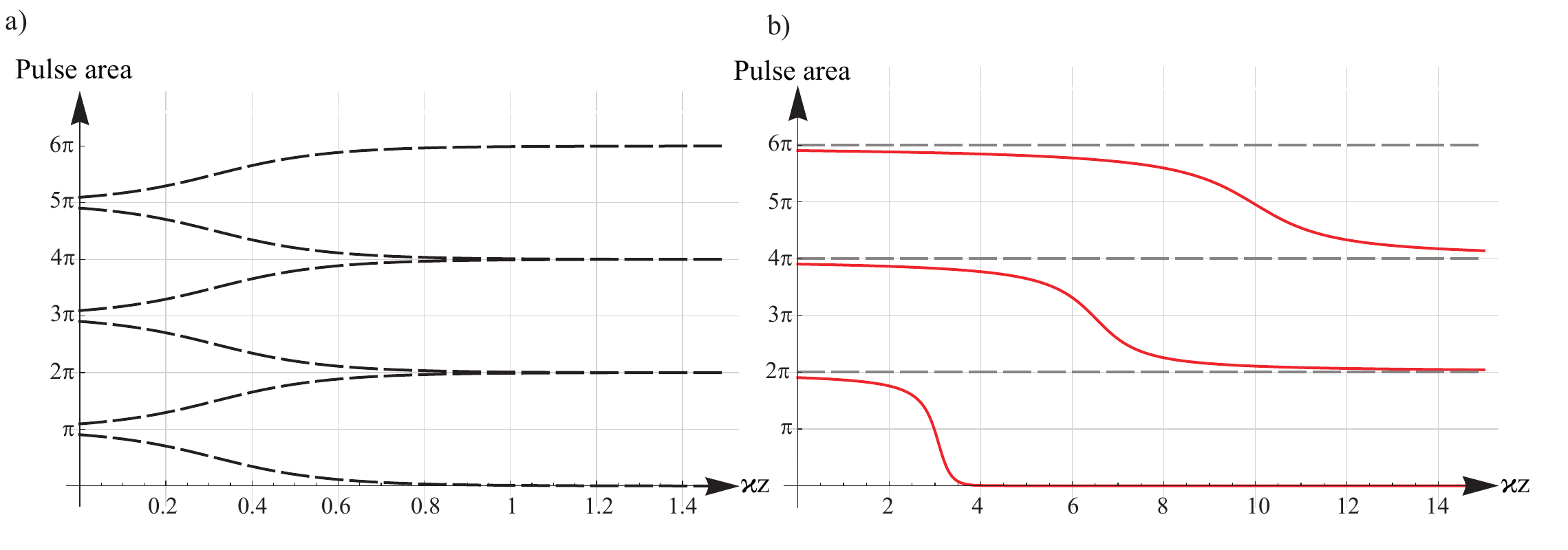}
\caption{ \textbf{a)} Pulse area  $\Theta_{p}(z)$ of McCall-Hahn area theorem \cite{McCall1969} versus optical depth   $\varkappa z$  for different input pulse areas ($\Theta_{p}(0): \pi \pm 0.1, 3\pi \pm 0.1, 5\pi \pm 0.1$);  \textbf{b)} pulse area $\Theta_{p}(z)$  of WPA theorem \eqref{Solution_Area_theor_inten_1} versus optical depth $\varkappa z$ for different input pulse areas ($\Theta_{p}(0): 2\pi-0.2, 4\pi-0.2, 6\pi-0.2$); the optical densities of atomic ensembles in the waveguide and in free space are assumed to be equal.}
\label{MC-Hahn_and_W-theorems}
\end{figure*}

The solution of Eq. \eqref{Envelope_area} strongly depends on the transverse properties of the waveguide modes. 
In the simplest case of a plane wave, where the field amplitude is spatially homogeneous in the cross-section ($f(\mathbf{r}_{\perp})=f_h(\mathbf{r}_{\perp})= 1$), we get an atomic function containing just the sum of the known terms corresponding to various dipole moments
(see also \cite{Lamb1971}): 

\fla{
\Phi(\theta_{p}) =
\sum_{m=1}^{M}\frac{S\xi_m \Omega_{m}}{2}\sin\big( \Omega_{m}\theta_{p}\big). 
\label{MC_function}
}
\noindent
For $M=1$ the derived Eq. \eqref{F_function} is reduced to the well-known McCall-Hahn pulse area theorem for the light pulse propagation in free space \cite{McCall1969,allen1975}.
In the limit of large optical depth the pulse area of the input pulse tends to $\Theta(\varkappa_m z\gg1)\rightarrow 2n \pi$ for the initial pulse area of $(2n-1)\pi<\Theta(0)<(2n+1)\pi$  as it is shown in Fig. \ref{MC-Hahn_and_W-theorems} a). 
Here, $\varkappa_m=\pi a^2\Omega_{m}^2 \xi_m = \frac{ \rho_m\omega}{4v_g \varepsilon_0 \varepsilon \hbar} |\langle \mathbf{d}_m\cdot \mathbf{e} \rangle|^2$ is a resonant absorption coefficient of $m$-th atomic group and $S=\pi a^2$ with $a$ being a radius of the waveguide.

For the transverse light mode with a Gaussian membrane function
\fla{
f(\mathbf{r}_{\perp} )=f_{g}(\mathbf{r}_{\perp})= \exp\{-\frac{(x^2+y^2)}{2a^2}\}.
\label{Memb_function_G}
} 
the integration over the cross-section in Eq. \eqref{F_function}  gives the following equation on the envelope area

\fla{
\big(\frac{\partial}{\partial z}+\frac{\gamma_w}{2v_g}\big)\theta_{p}=
-\sum_{m=1}^{M}\frac{\varkappa_m}{\Omega_{m}}
\frac{\sin^2{\big(\Omega_{m}\theta_{p}/2\big)}}{\Omega_{m}\theta_{p}/2}.
\label{W-area_theor}
}

\noindent
For Gaussian mode WPA theorem shows that $\frac{\partial}{\partial z}\theta_{p}(z)<0$ for $2n\pi<\Theta_{m,p}<2(n+1)\pi$ with $n\in \{0,1,2,...\}$ and
$\Theta_{m,p}=\Omega_{m}\theta_{p}$ being conventional pulse area of the $m$-th resonant atoms without spatial dependence as if the $m$-th atoms were located in the center of the waveguide.  
At the same time, $\frac{\partial}{\partial z}\theta_{p}(z)=0$ for
$\Theta_{m,p}=2n\pi$ and negligibly weak non-resonant losses ($\frac{\gamma_w L}{2v_g}\ll 1$). 
In other words, if the initial pulse area $\Theta_{m,p}(0)$ is less than $2\pi$, at the output the pulse area decreases to zero. 
This fact distinguishes the WPA theorem from the McCall-Hahn theorem \cite{McCall1969},  where $2\pi$ pulse is formed even if $\Theta_{m,p}(0)>\pi$. 

Similarly as in plane-wave case, in the limit of small pulse area the Eq. \eqref{W-area_theor} is reduced  to the linearized Lambert-Beer equation:
\fla{
\frac{\partial}{\partial z}\theta_{s}(z)=
-\frac{1}{2}\big(\frac{\gamma_w}{v_g}+\sum_{m=1}^{M}\varkappa_m\big)
\theta_{s}(z),
\label{Area_theor_weak_s}
}
\noindent
with exponential solution
\fla{
\theta_{s}(z)=\theta_{s}(0)e^{-\alpha z/2},
\label{Solution_Area_theor_weak_s}
}

\noindent
where $\alpha=\frac{\gamma_w}{v_g}+\sum_{m=1}^{M}\varkappa_m$ is a total absorption coefficient. In the case of only a single atomic type, i.e. $M=1$ and $\Theta_{m,p}(z)=\Theta_{p}(z)$, the Eq. \eqref{W-area_theor} is simplified to 
\fla{
\big(\frac{\partial}{\partial z}+\frac{\gamma_w}{2v_g}\big)\Theta_{p}=
-\varkappa_1
\frac{\sin^2{\big(\Theta_{p}/2\big)}}{\Theta_{p}/2}.
\label{W-area_theor_1}
}

\subsection{Gaussian-type modes}
The other remarkable example is application of WPA theorem to the  interaction of light pulse with two-level atoms in typical single mode optical fiber.
The membrane function of a quasi-linearly polarized mode in the single-mode fiber is often described by a zero-order Bessel function \cite{okamoto2021fundamentals}. 
Here, we consider two possible cases for filling the fiber with two-level atoms. 
In the first case ($f_{1}(\mathbf{r}_{\perp} )$), the atoms are evenly distributed inside the fiber core, and in the second case ($f_{2}(\mathbf{r}_{\perp} )$), they are also located in the fiber cladding, filling the entire fiber volume evenly.
The membrane functions of light modes in these two cases are:

\fla{
f_{1}(\mathbf{r}_{\perp} )&=\begin{cases}
J_0(u r_{\perp}/ a) &  r_{\perp} \le a \\
0 & r_{\perp}>a
\end{cases},
\\
f_{2}(\mathbf{r}_{\perp} )&=\begin{cases}
J_0(u r_{\perp}/ a),  & r_{\perp} \le a \\
\frac{J_0(u)}{K_0 (w)} K_0 (w r_{\perp}/a) & r_{\perp}>a
\end{cases},
}
where $J_0(x)$ is Bessel function of the first kind, $K_0(x)$ is modified Bessel function of the second kind, $u$ is the normalized transverse phase constant, $w$ is normalized transverse attenuation constant (propagation constant), $a$ is a fiber core radius.     

\begin{figure}
\includegraphics[width=0.45\textwidth]{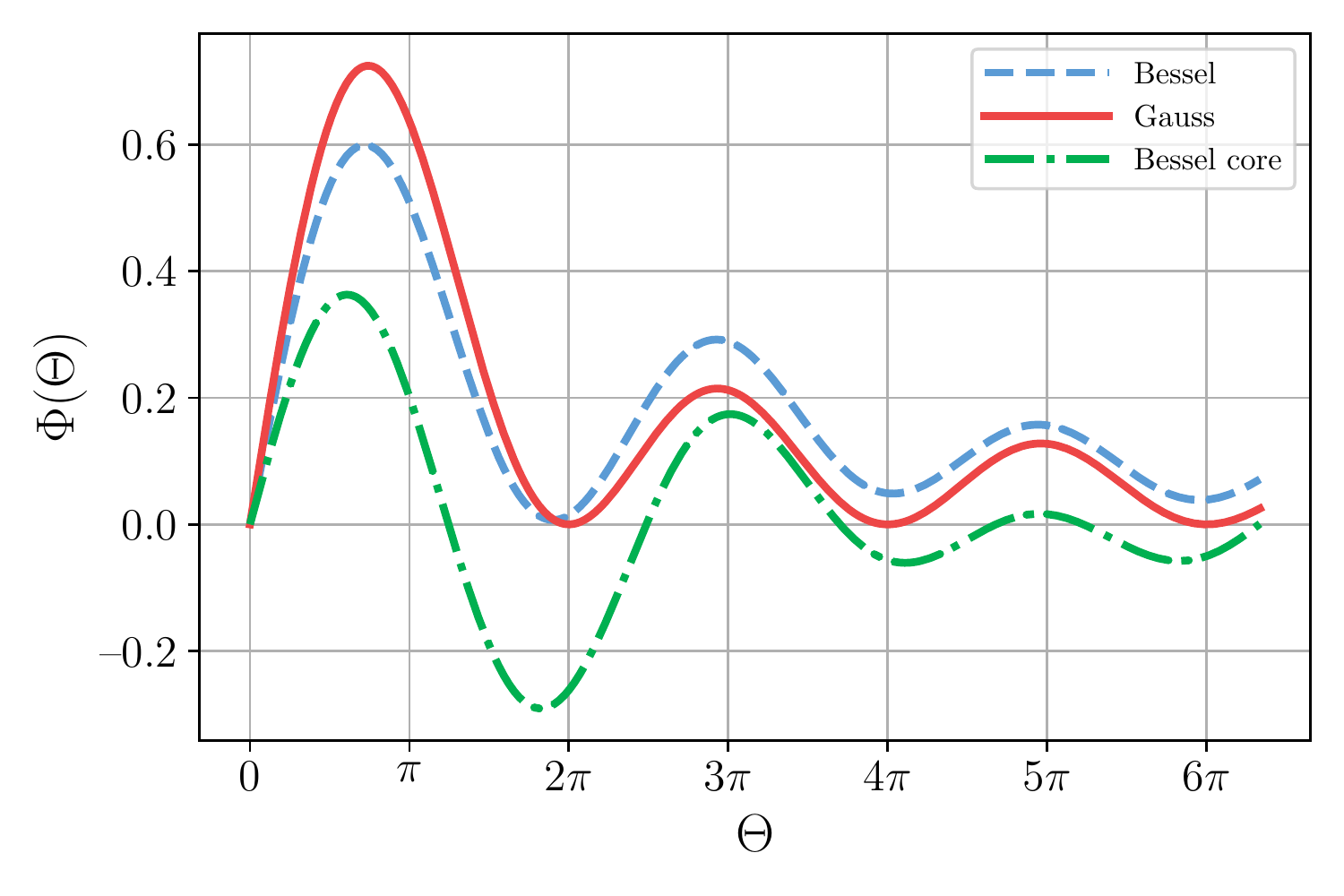} 
\caption{The atomic functions $\Phi_{g,1,2}(\Theta)$ for Gaussian, Bessel and Bessel core membrane functions $f_{g,1,2}(\mathbf{r}_{\perp})$. }
\label{Membran_functions}
\end{figure}

We compare properties of these Gaussian-type modes to the derived Gaussian mode assuming presence of uniformly distributed single atomic specie.
We numerically integrate the right-hand side of Eq.\eqref{F_function}
for $f_{1,2}(\mathbf{r}_{\perp} )$ and $f_{g}(\mathbf{r}_{\perp} )$.
The resulted functions $\frac{\Omega_1}{\varkappa_{1}}\Phi_{g,1,2}(\theta_{p})$ describe the contributions of the total polarization in the equation for given envelope area as it is in Fig. \ref{Membran_functions}.
The Gaussian and Bessel modes give very similar results for their atomic functions  $\Phi_{g}(\theta_{p})$ and $\Phi_{2}(\theta_{p})$, which are greater than zero for any $z$.
A significantly noticeable difference between these two functions occurs only starting from the second minimum at $\Theta = 4\pi$, where the function $\Phi_{2}(\theta_{p})$ becomes nonzero.
When atoms are uniformly distributed only inside the fiber core, the function $\Phi_{1}(\theta_{p})$ has the form of a damped oscillatory function.
Similarly to the  McCall-Hahn area theorem, the branching points of the solution also appear at $\Theta_{p}\approx 1.25\pi$, $\Theta_{p}\approx 3.75\pi$, ..., where the solution of pulse area $\Theta_{p}(z)$ increases from the branching points to the values of $2.5\pi$ and $4.5\pi$.
The similar quantitative and qualitative behaviors of  $\Phi_{g}(\theta_{p})$ and $\Phi_{2} (\theta_{p})$ allow to use the analytical solution of the Gaussian mode for description of the both cases at $\Theta_p(0) < 4\pi$. 

\subsection{Evolution in optically dense medium}
We study the evolution of the pulse area in a waveguide with Gaussian mode and optically dense medium. 
For single atomic group and negligibly weak non-resonant absorption  $\frac{\gamma_wz}{2v_g}\ll1$, the exact analytical solution of Eq.\eqref{W-area_theor_1} is

\fla{
&T_n(\Theta_{p}(z))=T_n(\Theta_{p}(0))-\frac{\varkappa_1}{2} z,
\label{Solution_Area_theor_inten_1}
}
where $T_n(\Theta_{p})$ is given by

\fla{
T_n(\Theta_{p})=\ln\big[\sin({\frac{\Theta_{p}}{2}})\big]-\frac{\Theta_p}{2}\cot{\frac{\Theta_{p}}{2}}.
\label{Solution_Area_theor_inten_2}
}

\noindent
Analytical solution of Eq. (\ref{Solution_Area_theor_inten_1})- (\ref{Solution_Area_theor_inten_2}) clearly reveals distinctive features of pulse area evolution in the optically dense single mode optical waveguides.
Figs. \ref{MC-Hahn_and_W-theorems} a), b) shows the behavior of the pulse areas for the McCall-Hahn area theorem and for the WPA theorem, respectively.
According to McCall-Hahn theorem \cite{McCall1969}, $2\pi$-pulse are generated for initial pulse area $\pi<\Theta_{p}(0)<2\pi$ as shown in Figs. \ref{MC-Hahn_and_W-theorems} a). 
In contract, WPA theorem states that formation of $2\pi$-pulse in the single mode waveguide is possible only for $\Theta_{p}(0)>2\pi n$ with $n\in\{1,2,...\}$.

The former fact is simply explained by asymptotical behaviour of the pulse area.
For the initial pulse area within a range $2n\pi<\Theta_{p}(0)<2(n+1)\pi$ the derivative becomes negative $\frac{\partial}{\partial z}\Theta_{p}(z)<0$ that results in asymptote  $T_{0}(\Theta_{p}(z\gg \varkappa_1^{-1}))\rightarrow \ln\big(\Theta_{p}(z)/2)\big)$.
According to Eq. \eqref{Solution_Area_theor_inten_2}, for $n=0$ the pulse area evolves asymptotically
\fla{
\Theta_{p}(z\gg \varkappa_1^{-1}) \cong e^{T_0(\Theta_{p}(0))-\varkappa_1 z/2}  \rightarrow 0,
}
while for $n=1,2,...$
\fla{
\Theta_{p}(z \gg \varkappa_1^{-1}) \cong  2\pi n \big( 1+\frac{1}{\frac{\varkappa_1 z}{2}-T_n(\Theta_{p}(0))}\big) \rightarrow  2\pi n.
\label{asymptotic}
}
\noindent
Remarkably, the asymptote of $\Theta_{p}(z\gg \varkappa^{-1})$ for $n\neq 0$ does not have an exponential character and slowly converges compared to the solution $\Theta_{p}(z\gg \varkappa_1^{-1})$ for $n=0$. Similarly, the formation of $2\pi$-pulse in the waveguide occurs at almost tenfold larger optical depth than for plane wave as depicted in Figs. \ref{MC-Hahn_and_W-theorems}a and  \ref{MC-Hahn_and_W-theorems}b.  The formation of the second and subsequent $2\pi$ pulses in the waveguide is also shifted into region of larger optical depth. 
One reason for that is the weaker interaction between the waveguide mode and atoms away from the center of the waveguide. 
In principle, this reduction may be compensated by an effective increase of the coupling constant via the decreased group velocity of the light modes $v_g\ll c $  (see  comments to \eqref{MC_function}) and stronger mode confinement in the waveguide. 

For the pulse areas close to  $2\pi n$ attenuation will be caused only by non-resonant losses with $\Theta(z)\approx 2\pi n \exp\{-\frac{\gamma_wz}{2v_g}\}$. Such pulses will persist the longest  soliton-like propagation for weak non-resonant losses ($\frac{\gamma_wz}{2v_g}\ll 1$).
However, at large deviations of the signal pulse area from $2 \pi n$ the contribution of resonant interaction in bringing the pulse area to 
$2\pi(n-1)$ becomes dominant.
Such an evolution is to be repeated until the light field is completely absorbed.



The presence of several different dipole moments $\mathbf{d}_m$ and, accordingly, different pulse areas  $\Theta_{m,p}$ significantly change the behavior of the pulse area. 
The formation of relatively stable  $2\pi$ pulses becomes possible, probably, only if the condition $\Theta_{m,p}=2n\pi$ is met for each atomic group $m$. 
A detailed study of the implementation of such scenarios for the formation of $2\pi$ pulses is beyond the scope of the current analysis.



\section{Multipulse excitation and echo signal emission}

\subsection{Pulse area theorem of echo signal in single mode waveguide}

Here we apply the WPA theorem to the description of the photon echo in an ensemble of two-level atoms uniformly distributed in the volume of a single-mode waveguide.
We assume that the signal pulse is launched at time $t=0$ and it's interaction with two-level atoms is governed by WPA theorem that is described in Sec.\ref{sec::math_formalism}. 
After signal an additional control pulse with a delay of $\tau$ is applied to the medium. The delay is assumed to be much longer than the signal pulse duration ($\tau\gg\delta t$). 

The signal pulse affects on the initial atomic population
$w_{m,s} (\mathbf{r}_{\perp},z)$ that changes the initial conditions  in the equation for the envelope area of the control pulse. 
With the help of Eqs. of \eqref{eq::Asclass}-\eqref{eq::w} the Eq. \eqref{Envelope_area} is modified accordingly to reflect the initial atomic population
\fla{
&\big(\frac{\partial}{\partial z}+\frac{\gamma_w}{2v_g}\big)\theta_{1}= \sum_{m=1}^{M}\frac{\xi_m}{2}\cdot
\nonumber
\\
&\int_S dxdy\Omega_{0,m} (\mathbf{r}_{\perp} )
w_{m,s} (\mathbf{r}_{\perp},z)
\sin\big( \Theta_{m,1}(\mathbf{r}_{\perp},z)\big),
\label{Envelope_area_с2}
}
where $w_{m,s} (\mathbf{r}_{\perp},z)=-\cos\big( \Theta_{m,s}(\mathbf{r}_{\perp},z)\big)$,  
$\Theta_{m,s}(\mathbf{r}_{\perp},z)=
\Omega_{0,m} (\mathbf{r}_{\perp} )\theta_{s}(z)$
is the pulse area of the signal pulse for an atom located at the point ($\mathbf{r}_{\perp},z$). The pulse area of signal can be expressed by previously derived Eqs. \eqref{Solution_Area_theor_inten_1}
for Gaussian-type membrane function.

Next, we derive an equation for the envelope area $\theta_e(z)$ of the simple two-pulse (primary) echo. 
Following the same procedure as with derivation of WPA theorem we integrate the equation for the field amplitude
over time. 
The echo is expected at $t=2\tau$, thus the integration is reasonable to start from the moment of time before the  echo at $t=3\tau/2$ and finish after echo irradiation  $t=5\tau/2$:
$\theta_{e}(z)=\int_{3\tau/2}^{5\tau/2} dt a_{e}(z,t)$. 
Since the echo signal is time-separated from other light pulses, we carry out the integration as in Sec. \ref{sec::math_formalism} and take into account the initial atomic polarization that results in the following equation for the envelope area  $\theta_{e}$ of the echo pulse:

\fla{
&\big(\frac{\partial}{\partial z}+\frac{\gamma_w}{2v_g}\big)\theta_{e}= \sum_{m=1}^{M}\frac{\xi_m}{2}\int_S dxdy\Omega_{0,m} (\mathbf{r}_{\perp} )\cdot
\nonumber
\\
&\big\{2\Gamma(\mathbf{r},\tau,...)P_m(\mathbf{r}_{\perp},z)\cos^2\big(\frac{1}{2} \Theta_{m,e}(\mathbf{r}_{\perp},z)\big)+
\nonumber
\\
&w_m(\mathbf{r}_{\perp},z)
\sin\big( \Theta_{m,e}(\mathbf{r}_{\perp},z)\big)\big\},
\label{Area_theor_e}
}

\noindent
where $\Theta_{m,e}(\mathbf{r}_{\perp},z)=
\Omega_{0,m} (\mathbf{r}_{\perp} )\theta_{e}(z)$, 
$P_m(\mathbf{r}_{\perp},z)$ is the source of the echo being the phasing part of the polarization of the m-th atomic ensemble taken at the central frequency, $w_m(\mathbf{r}_{\perp},z)$ is the non-oscillating part of the population of m-th atomic ensemble at the central frequency \cite{Moiseev1987,Urmancheev2019,Moiseev2020}. 
Note that in the case of plane waves $f(\mathbf{r}_{\perp})= 1$ and $\gamma_w=0$, the analytical solution of Eq.\eqref{Area_theor_e} for $M=1$ is found using substitution $\theta_e(z)=\frac{2}{\Omega_{0}}\arctan u(z)$. 
This solution shows that the pulse area of each echo signal $\Theta_e=\Omega_{0}\theta_e(z)$ does not exceed $\pi$ and decreases to zero at high optical depth \cite{Moiseev2020}.

The term $\Gamma(\mathbf{r},\tau,...)$  describes the phase relaxation in the interval between signal pulse and echo pulse. As previously, the influence of phase relaxation during the relatively short interaction time between light pulses and the polarization is neglected for a simple integration \cite{Moiseev2020}. 
At the same time, the phase relaxation in the interval between the pulses and the echo pulse is included. 
For sake of simplicity, the exponential relaxation factors $\Gamma(...) =e^{-(2 \tau/T_M)^x}$ for primary echo and $e^{-(2 \tau/T_M)^x}$ for ROSE-signal are used \cite{Mims1968}, where $x$ is the fitting power factor.
It should be emphasised that various decoherence mechanisms, particularly depending on the level of atom excitation, may require specific study. One such mechanism is the effect of instantaneous spectrum diffusion \cite {Cone1988,2014-LP-Thiel}.

\subsection{Pulse area of two-pulse (primary) photon echo and all subsequent echoes}
Below we analyse the solution of Eq. \eqref{Area_theor_e} for the two-pulse (primary) echo and the pulse area of all subsequently induced echoes in an optically dense two-level medium under the assumption of weak phase relaxation ($\Gamma(\mathbf{r},\tau,T_M,\Theta_{s,1},\beta)=1$). 
\begin{figure}
\includegraphics[width=\linewidth]{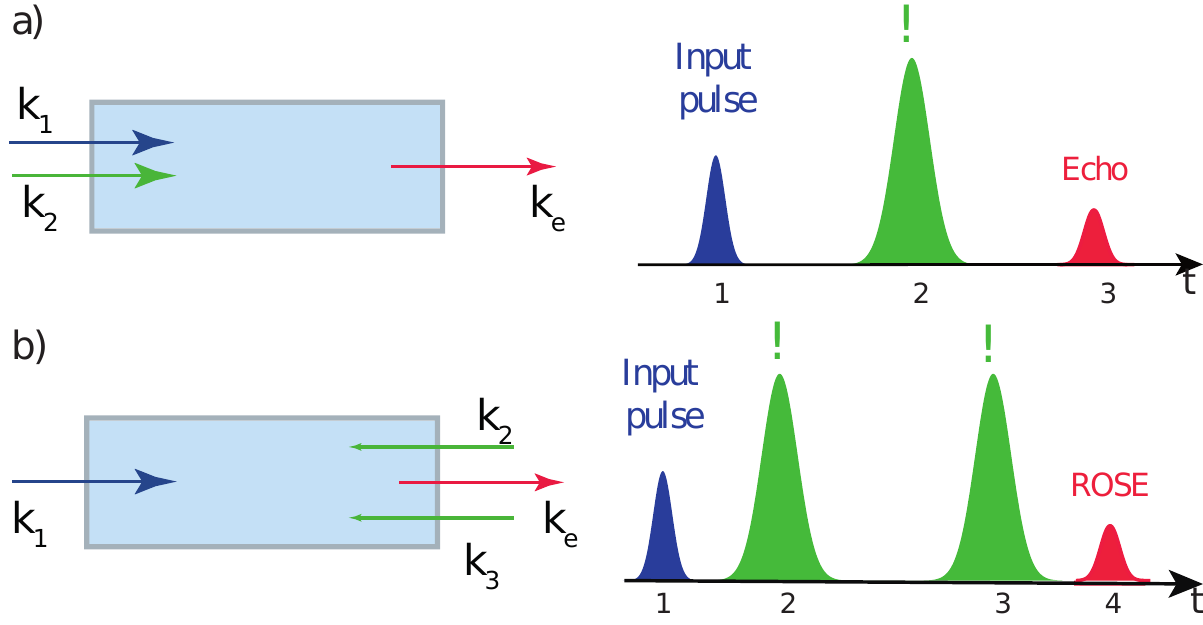}
\caption{Spatial schemes and temporal sequences of light pulses in \textbf{a)} the primary echo, \textbf{b)} the revival of silenced echo (ROSE) protocol.}
\label{W-sequence} 
\end{figure}
First, we consider the case of two co-propagating  pulses as illustrated in Fig. \ref{W-sequence} a).
The initial conditions of the envelope equation contain the components of atomic inversion  $w_m(\mathbf{r}_{\perp},z)=w_m^{pe}(\mathbf{r}_{\perp},z)$ and  coherence $P_m(\mathbf{r}_{\perp},z)=P_m^{pe}(\mathbf{r}_{\perp},z)$, which are expressed as

\fla{
\noindent
&w_m^{pe}(\mathbf{r}_{\perp},z)=-
\cos\big( \Theta_{m,s}(\mathbf{r}_{\perp},z)\big)
\cos\big( \Theta_{m,1}(\mathbf{r}_{\perp},z)\big),
\nonumber
\\
& P_m^{pe}(\mathbf{r}_{\perp},z)=
\sin\Theta_{m,s}(\mathbf{r}_{\perp},z)
\sin^2\big(\frac{1}{2} \Theta_{m,1}(\mathbf{r}_{\perp},z)\big). \label{parameters_1}
}
\noindent

The only difference between the expressions in Eqs.  \eqref{parameters_1} and similar solutions for a plane wave \cite{Moiseev1987,Moiseev2020} is the dependence of the pulse areas on the transverse coordinate $\mathbf{r}_{\perp}$.
We integrate the atomic response along the transverse plane of the waveguide in the same way as in Eq. \eqref{W-area_theor}.
The resulted equation for the envelope area of echo pulse is
\fla{
\big(\frac{\partial}{\partial z}+\frac{\gamma_w}{2v_g}\big)\theta_{e}=
&\sum_{m=1}^{M}\frac{\varkappa_m}{\Omega_{m}}
\{\frac{\Gamma(\tau,T_M,...)}{2} 
I_{s,m}(\theta_{s},\theta_{1}, \theta_{e})
\nonumber\\
&-S_m(\theta_{e};\theta_{s};\theta_{1})
\},
\label{Envelope_echo-area-a}
}

where
\fla{
I_{s,m}(\theta_{s},\theta_{1}, \theta_{e})=&
\frac{\sin^2{\big(\Omega_{m}\theta_{s}/2\big)}}{\Omega_{m}\theta_{s}/2}
+S_m(\theta_{s};\theta_{e})
\nonumber
\\
&-S_m(\theta_{s};\theta_{1})-
S_m(\theta_{s};\theta_{1};\theta_{e}),
}

\begin{figure*}
\centering
\includegraphics[width=\linewidth]{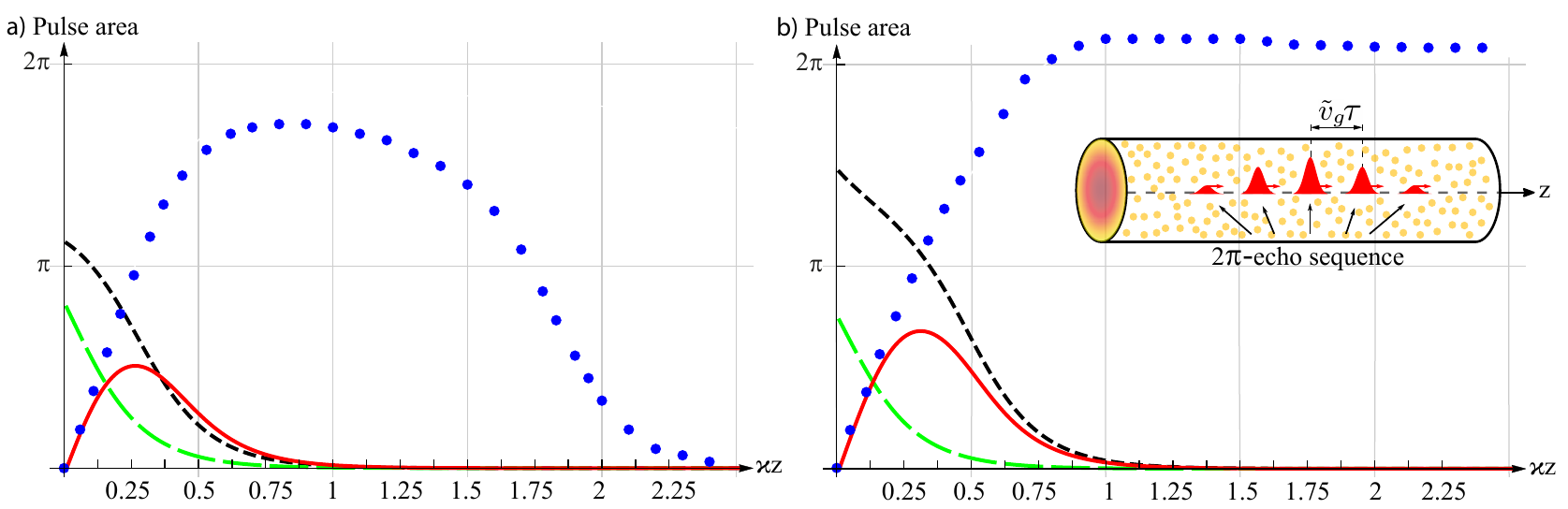}
\caption{ 
Pulse areas for single atomic group $M=1$ of signal pulse $\Theta_s(z)$ (\textcolor{green}{green} dashed line), first control pulse $\Theta_1(z)$ (\textbf{black} short dashed line), primary echo $\Theta_e( z)$ (\textcolor{red}{red} solid line) and all echo pulses $\Theta_{\Sigma, e}(z)$ (\textcolor{blue}{blue} dotted line).
\textbf{a)} $\Theta_s(0)+\Theta_1(0)<2\pi$
\textbf{b)} $2\pi<\Theta_s(0)+\Theta_1(0)<4\pi$.
Inset conceptually illustrates formation of an $2\pi$-echo sequence in the waveguide at some moment of time with total pulse area $\Theta_{\Sigma, e}=2\pi$, where  $\tau$ is a time interval between the two exciting pulses, $\tilde{v}_g$ is a group velocity  weakened by the light interaction with two-level atoms.}
\label{W-area_theorem_new}
\end{figure*}

\fla{
&S_m(\theta_{p};\theta_q)=
\nonumber
\\
&\frac{1}{2}\sum_{n=1}^{2}\frac{\sin^2{\big(\Omega_{m}(\theta_{p}+(-1)^n\theta_{q})/2\big)}}
{\Omega_{m}(\theta_{p}+(-1)^n\theta_{q})/2},
\\
&S_m(\theta_{p};\theta_q;\theta_r)=
\nonumber
\\
&\frac{1}{4}\sum_{n=1}^{2}\sum_{l=1}^{2}\frac{\sin^2{\big(\Omega_{m}(\theta_{p}+(-1)^n\theta_{q}+(-1)^l\theta_{r})/2\big)}}
{\Omega_{m}(\theta_{p}+(-1)^n\theta_{q}+(-1)^l\theta_{r})/2}.
\label{S-functions}
}
For a single dipole moment type the resulted pulse areas of the input signal pulses $\Theta_{s}(z)$, control pulse $\Theta_{1}(z)$, primary echo $\Theta_{e}(z)$ and the total pulse area of all echoes $\Theta_{\Sigma,e}$ are depicted in Fig. \ref{W-area_theorem_new}. 
As seen in the Figs.\ref{W-area_theorem_new} a), b), the pulse area of the primary echo pulse $\Theta_e(z)$ experiences growth reaching its maximum and then decreasing to zero.
The pulse area of the all echoes $\Theta_{\Sigma,e}(z)$ is defined as the difference between the total pulse area $\Theta_{\Sigma}(z)$ and individual  pulse areas of the signal and control pulses $\Theta_{s}(z)$, $\Theta_{1}(z)$ as in \cite{HAHN1971}, where the evolution of total pulse area $\Theta_{\Sigma}(z)$ is described by Eq. \eqref{W-area_theor_1} with initial pulse area  $\Theta_{\Sigma}(0)=\Theta_{s}(0)+\Theta_{1}(0)$.

Fig. \ref{W-area_theorem_new} a) shows that if the initial total area $\Theta_{\Sigma}(0)$ is less than $2\pi$, then the pulse area of all echo signals $\Theta_{\Sigma,e}(z)$ tends to zero at large optical depth $ z\gg 1/\varkappa$. 
In contrast, if $\Theta_{\Sigma}(0)>2 \pi$, then the pulse area of all echo signals $\Theta_{\Sigma,e}(z\gg 1/\varkappa)$ tends to $2\pi$, as it is seen in Fig. \ref{W-area_theorem_new} b).
This behaviour does not coincide with the prediction of McCall-Hahn area theorem, where the threshold of the total area is $\pi$ for the $2\pi$ pulse formation (see Fig. \ref{MC-Hahn_and_W-theorems} a)).
It is also seen in Fig. \ref{W-area_theorem_new} a), b), that in both the cases the primary echo amplitude increases with the optical density approaching its maximum  without  reaching to $2 \pi$. 
At large value of optical depth the primary echo gradually decays to zero.
The growth of the pulse area to its maximum and its subsequent attenuation to zero is valid for each of the subsequent echo, which is a common property of WPA theorem and free space theory \cite{Moiseev1987,Urmancheev2019}.
The total pulse area is the sum of the pulse areas of several echo signals generated sequentially in a medium, since total pulse area of all pulses tends to $2\pi$, if $\Theta_{\Sigma}(0)>2 \pi$.
Thus, the sequence of generated echoes in a single mode waveguide  behaves as a single $2\pi$-pulse, as in case with plane waves \cite{Moiseev2020} and we indicate it  in Fig. \ref{W-area_theorem_new} b) as a $2\pi$-echo sequence
(see also discussion in Appendix A about the $2\pi$-pulse excitation in a single mode waveguide).

Following the approach \cite{Moiseev2020,Moiseev2022} and using one-dimensional Eq. \eqref{Area_theor_e} with corresponding expressions for the phasing polarization and initial atomic inversion, it is also possible to get equations for an envelope areas of an arbitrary echo signal,  which can be easily solved numerically.
Below, we  developed photon echo WPA theorem to the photon echo in ROSE protocol \cite{Damon2011,Optica-Liu:20,2022-Minnegaliev-JEPTL}. 

\section{ROSE protocol in a laser-written waveguide crystal}

\subsection{Waveguide pulse area equation of ROSE signal}

ROSE is echo protocol for quantum storage being closest to the convectional two-pulse echo. If implemented properly, it suppresses unwanted quantum noise that appears at primary echo emission. The spatial and temporal schemes of ROSE protocol are shown in Fig. \ref{W-sequence}b).
In ROSE protocol a two-level medium is firstly excited by a weak signal pulse at time $t=0$. Two $\pi$ pulses are applied to medium at $t=\tau$ and $t=3\tau$ that forms an echo at $t=4\tau$. 
The control pulses should propagate in the opposite direction with respect to the direction of the signal pulse to mitigate generation of the primary photon echo.

For signal, echo, first and second $\pi$ pulses we use indexes $(s,e,1,2)$, respectively.
The pulse area after the weak signal and first $\pi$ pulse are described by Eq. \eqref{Solution_Area_theor_inten_1}, \eqref{Solution_Area_theor_inten_2}
(where to make replacements: $z \rightarrow(L-z)$ and $\theta_1(0) \rightarrow \theta_1(L)$),
due to negligibly small pulse area of the signal. 
Since the first control pulse has a pulse area close to $\pi$ it strongly affects the propagation of the second control pulse ($p=2$).
In this case, similarly to \cite{HAHN1971}, we get  the following equation for the envelope area  $\theta_{2}$ of the second control pulse:

\fla{
&\big(-\frac{\partial}{\partial z}+\frac{\gamma_w}{2v_g}\big)\theta_{2}= -\sum_{m=1}^{M}\frac{\xi_m}{2}\cdot
\nonumber
\\
&\int_S dxdy\Omega_{0,m} (\mathbf{r}_{\perp} )
\cos\big( \Theta_{m,1}(\mathbf{r}_{\perp},z)\big)
\sin\big( \Theta_{m,2}(\mathbf{r}_{\perp},z)\big).
\label{Envelope_area_с2n}
}

\noindent
By using Gaussian membrane function in Eq.\eqref{Envelope_area_с2n} and performing the calculation similar to the derivation of Eqs. \eqref{W-area_theor} and \eqref{Envelope_echo-area-a}, we get:

\fla{
\big(-\frac{\partial}{\partial z}+\frac{\gamma_w}{2v_g}\big)\theta_{2}=
-\sum_{m=1}^{M}\frac{\varkappa_m}{\Omega_{m}}S_m(\theta_{2};\theta_{1}),
\label{Envelope_area_c2m}
}

\noindent
where $S_m(\theta_{p};\theta_{q})$ is given in Eq.\eqref{S-functions}.
For the particular case of weak second control field ($\Omega_{m}\theta_{2}\ll\pi$),  Eq. \eqref{Envelope_area_c2m} transforms into the linear equation for $\theta_{2}$:

\fla{
\big(-\frac{\partial}{\partial z}+\frac{\gamma_w}{2v_g}\big)\theta_{2}(z)=
-\frac{1}{2}\sum_{m=1}^{M} \varkappa_m(\Theta_{m,1}(z))
\theta_{2},
\label{Area_theor_c2m2}
}

\noindent
where  $\varkappa_m(\Theta_{m,1} (z))$  is a changed absorption coefficient of $m$-th atomic group ($\Theta_{m,1}(z)=\Omega_{m}\theta_{1}(z)$):

\fla{
\varkappa_m(\Theta_{m,1}(z))=\varkappa_m
\big\{2
\frac{\sin{\big(\Theta_{m,1}(z)\big)}}{\Theta_{m,1}(z)}
-\frac{\sin^2{\big(\Theta_{m,1}(z))/2\big)}}{(\Theta_{m,1}(z)/2)^2}
\big\},
\label{formula_5}
}

\noindent
where $\varkappa_m(\Theta_{m,1}\rightarrow 0)=\varkappa_m$ and $\varkappa_m(\Theta_{m,1})<0$ for $\tan(\Theta_{m,1}/2)>\Theta_{m,1}$, i.e. for $\Theta_{m,1}>2.3311$. In this condition the second control pulse is amplified, if non-resonant losses are weak $\gamma_w L \ll 1$.
If $\Theta_2(0)<2\pi$, the amplification of second pulse $\Theta_2(z)$ is replaced by its attenuation to zero at large optical depth $z\gg 1/\varkappa$. 
As in the case of the primary echo, the pulse area of ROSE protocol $\Theta_e(z)$ experiences growth to a maximum and decrease to zero at large value of optical depth. 
The evolution of the envelope area $\theta_e(z)$ in ROSE sequence is described by Eq. \eqref{Area_theor_e}, where the initial atomic population  $w_m(\mathbf{r}_{\perp},z)=w_m^{re}(\mathbf{r}_{\perp},z)$ and phasing coherence $P_m(\mathbf{r}_{\perp},z)=P_m^{re}(\mathbf{r}_{\perp},z)$ are \cite{2021-PRB-Minnegaliev}:

\fla{
 w_m^{re}(\mathbf{r}_{\perp},z)=
 & \cos\big( \Theta_{m,s}(\mathbf{r}_{\perp},z)\big)\prod_{p=1}^{2} \cos\big( \Theta_{m,p}(\mathbf{r}_{\perp},z)\big)
\nonumber
\\ 
 \cong &-\cos\big( \Theta_{m,1}(\mathbf{r}_{\perp},z)\big)\cos\big( \Theta_{m,2}(\mathbf{r}_{\perp},z)\big),
\nonumber
\\
 P_m^{re}(\mathbf{r}_{\perp},z)
\cong &
\Omega_{0,m} (\mathbf{r}_{\perp} )\theta_{s}(0)e^{-\alpha z/2}
\prod _{p=1}^{2}
\sin^2\big(\frac{1}{2} \Theta_{m,p}(\mathbf{r}_{\perp},z)\big),
\label{parameters_2}
}

\noindent
where $\cos\big( \Theta_{m,s}(\mathbf{r}_{\perp},z)\cong 1$ and  
\fla{
\sin\Theta_{m,s}(\mathbf{r}_{\perp},z)\cong \Theta_{m,s}(\mathbf{r}_{\perp},z)=\Omega_{0,m} (\mathbf{r}_{\perp} )\theta_{s}(0)e^{-\alpha z/2}, 
\label{pulse_area_s}
}
with pulse areas $\Theta_{m,p}(\mathbf{r}_{\perp},z)=\Omega_{0,m} (\mathbf{r}_{\perp} )\theta_{p}(z)$  (p=1,2)
being close to $\pi$ for highly efficient quantum memory ROSE protocol. 
In the studied counter-propagating geometry  $"\frac{\partial}{\partial z}"$ should be  replaced by $"-\frac{\partial}{\partial z}"$ in  Eqs. \eqref{W-area_theor}, \eqref{Envelope_area_с2},  when considering the second control pulse
(see Eqs. \eqref{Envelope_area_с2n}-\eqref{Area_theor_c2m2}).
At the same time, the efficiency of echo signal retrieval does not depend on the direction of its radiation due to the low optical density.

We note that $\theta_s, \theta_{1}, \theta_{2}$ in \eqref{Area_theor_e}, \eqref{parameters_1}, \eqref{parameters_2} can be found according to the Eqs. \eqref{W-area_theor} and \eqref{Envelope_area_с2} for the forward and backward propagation of the control pulses 
(with replacement $"\frac{\partial}{\partial z}"$  by  $"-\frac{\partial}{\partial z}"$).  
Similar as with Eqs.\eqref{Area_theor_c2m2}, \eqref{formula_5}, we find that the absorption coefficients $\varkappa_m(\Theta_{1}(z);\Theta_{2}(z))$ in \eqref{formula_4_dop} can be positive or negative, thereby leading to absorption or amplification of the echo signal.
The last case does is not suitable for QM due to quantum noise in the echo signal. 
For generality, in Appendix B, we provide equations for the pulse area of ROSE echo signal for arbitrary intensities of the signal and control fields in the waveguide with optically dense medium.

In our experiment, both control laser pulses have phase and amplitude modulations \cite{2021-PRB-Minnegaliev} that provide a uniform spectral and spatial excitation of atoms in the central part of the waveguide cross-section, allowing a significant increase in the efficiency of the echo's signal emission.

\subsection{Experimental setup}

\begin{figure*}
\centering
\includegraphics[width=\linewidth]{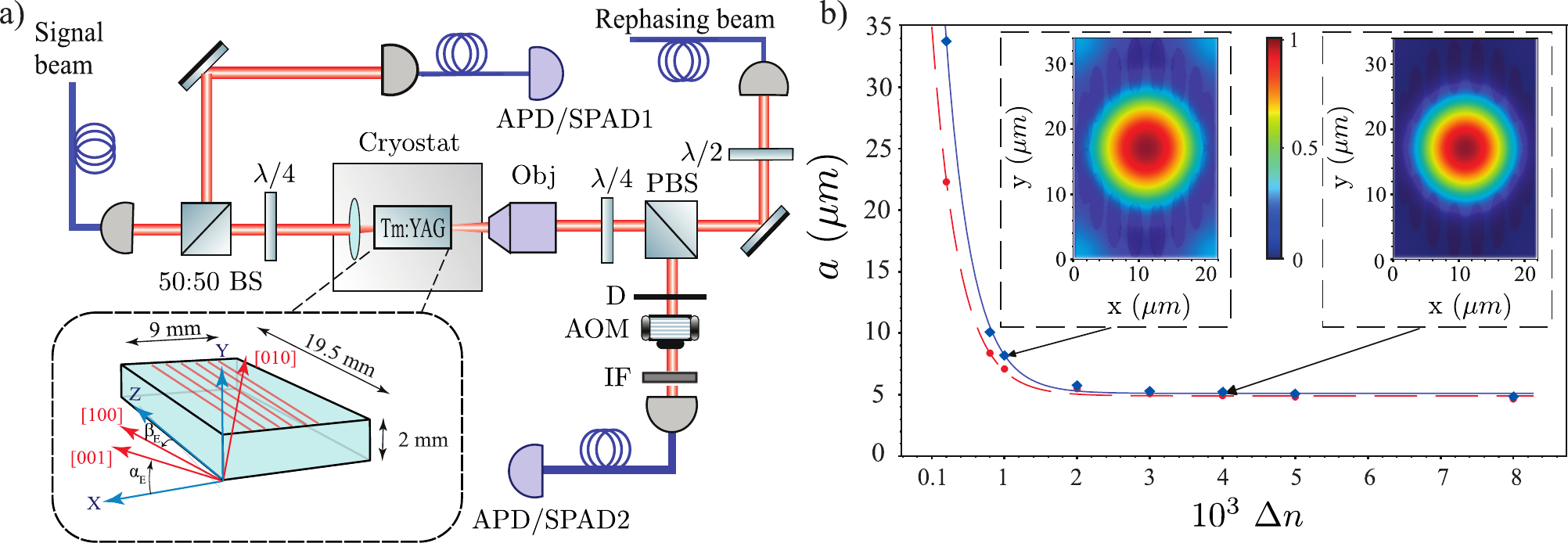}
\caption{ \textbf{a)} Simplified experimental scheme, where IF is an interference filter, Obj is an objective, also there is one more objective installed inside the cryostat on the left side of the crystal, D is a diaphragm, AOM is an acousto-optic modulator, BS is non-polarizing beam splitter, PBS is polarizing beam splitter, $\lambda/4$ is quarter wave plate,
$\lambda/2$ is half wave plate,
APD is an avalanche photodiode, SPAD is a single photon avalanche photodiode, $\alpha_E = 11.3^\circ$, $\beta_E = 4.3^\circ$  are conventional Euler rotation angles. 
\textbf{b)} 
Simulated full width half maximum $a$ of the laser-written waveguide eigen mode at different values of the refractive index contrast $10^{-4}<\Delta n<8\cdot10^{-3}$ between the elliptical tracks and the core. The \textcolor{red}{red} dashed and \textcolor{blue}{blue} solid curves correspond to widths along two orthogonal axes, $x$ and $y$, respectively. 
Two inserts show the color map of normalized spatial distribution of the eigen mode's intensity for $\Delta n=10^{-3}$ and $\Delta n=4\cdot 10^{-3}$.
}
\label{Fig::exp-setup}
\end{figure*}

We implement ROSE protocol in the waveguide that being laser-written in \tmYAG crystal with Tm$^{3+}$ doping of 0.01 $\%$ and dimensions 2x9x19.5 mm.
The crystal is one of the well-known crystals, convenient for testing and development of optical QM protocols \cite{2010-JofLum-CHANELIERE,2021-PRB-Minnegaliev}. 
Simplified experimental setup is depicted in Fig. \ref{Fig::exp-setup}a.
Laser light from a single-mode fiber is coupled into a single waveguide by an objective (Edmund Optics ELWD 10x 59877) on one side and an aspheric lens (Thorlabs A280TM-B) installed in a cryostat from another side of the crystal with   an efficiency of 30$\%$.
Light is detected by an avalanche photodetector (APD, Thorlabs APD120A/M) connected to an oscilloscope (Tektronix DPO7104C) or a single photon counting module (SPCM, Excelitas SPCM-AQRH).
The crystal is glued with silver paste to a cold finger and placed in a closed-cycle cryostat (Montana Instruments Corp.) with a temperature of 3.2$\pm$0.1 K.
Crystallographic axes of the crystal are oriented with respect to crystal edges, as it is shown in the inset of Fig. \ref{Fig::exp-setup} a), where $\alpha_E = 11.3^\circ$, $\beta_E = 4.3 ^\circ$, and $\gamma_E = 0^\circ$ are conventional Euler rotation angles.

The type-III depressed cladding single-mode waveguides were produced by a femtosecond laser writing technique in \tmYAG crystal \cite{Skryabin20}.
Each waveguide is composed of 18 elliptical tracks with axes of 2 and 8 $\mu$m around a core with diameter of 18 $\mu$m. 
The femto-second laser pulses change the refractive index of the elliptical tracks by $\sim 10^{-3}$ that confines the light within the inner core.
Propagation losses for vertical polarization  (parallel to 2 mm edge of the crystal) is 0.66 dB/cm, for horizontal polarization (parallel to 9 mm edge) is 1.13 dB/cm. 
Totally twenty one type-III single-mode waveguides have been written in the crystal along $Z$ axis. 
Since the exact difference in refractive index $\Delta n$ is unknown, we perform finite difference time-domain simulation of the eigenmodes for different values of $\Delta n$ \cite{codeYAG}. 
If $\Delta n > 2 \cdot 10^{-3}$ the  waveguide supports symmetrical Gaussian mode with half width at half maximum of $\sim 5.5$ $\mu$m for both axis as shown in Fig. \ref{Fig::exp-setup} b). At lower values of $\Delta n$ the mode becomes elliptical with the widths for orthogonal axes being different more than two times.
At the same time, the experimentally measured modes have symmetrical Gaussian character that witnesses the actual width of the waveguide mode to be $\sim 5.5$ $\mu$m.
Hence, further on we assume the width of $a=5.5$ $\mu$m in membrane function $f_{g}(\mathbf{r}_{\perp})$.

It should be noted that recent studies have conducted experiments on the implementation of photon echo AFC  protocol in waveguides fabricated  by the same technology \cite{Saglamyurek2011,Thiel_2012,Saglamyurek2015,Corrielli-PRApplied-2016, Optica-Liu:20}.
Whereas the cited publications used type I, type II, and type IV waveguides, we present type III waveguides in this work. The advantage of type III waveguide is a  support of arbitrary polarization in a waveguide mode.

The continuous wave Titan-Sapphire laser (Tekhnoscan TIS-SF-777) is tuned to \tmtrans optical transition of Tm$^{3+}$ ($\lambda$ $\approx$ 793.365 nm). 
Two acousto-optic modulators (AOMs) are used to sample microsecond-scale pulses for performing primary echo measurements and the ROSE protocol.  
The outputs of AOMs are spatially filtered by single mode fibers and sent parallel to the long edge of the crystal (19.5 mm) in counter-propagating geometry.
The linearly polarized signal is routed via 50:50 beam splitter to the APD for intensity monitoring as a reference and to the crystal, as shown in Fig.\ref{Fig::exp-setup}a. After passing through the quarter wave-plate the beams are circularly-polarized.

Tm$^{3+}$ ions substitute yttrium ions in six ($M=6$) crystallographically equivalent but orientationally inequivalent sites of \YAG crystal. 
The light beams with circular polarization interact with all these sites.
The Rabi frequency of optical transition $\Omega_{m}=E_0 \langle \mathbf{d}_m\cdot \mathbf{e} \rangle/\hbar$ is larger for the third and fourth site.
In order to make the pulse area of the control pulses to be close to $\pi$ for the atoms in wider spectral range, we use phase-modulated control laser pulses. 
The theoretical modeling of the control pulse Rabi frequencies and phase modulation is given by the relations (see also Appendix B):
\begin{equation}
 \Omega_m(\tau,\mathbf{r}_{\perp})=\Omega_{m,1}(\tau-t_1,\mathbf{r}_{\perp})+\Omega_{m,2}(\tau-t_2,\mathbf{r}_{\perp}), 
 \label{control_pulses}
\end{equation}

\noindent
where
\begin{align}
\Omega_{m,p}(\tau-t_1,\mathbf{r}_{\perp})&=
\dot{A}_{m,p} (\tau-t_1,\mathbf{r}_{\perp}) e^{-i B_p(\tau-t_p)},
\label{chirped-pulses-1}
\\
\dot{A}_{m,p}(t,\mathbf{r}_{\perp}) & = \frac{\chi_{m,p}(\mathbf{r}_{\perp})}{\pi \delta t_{p}} \frac{e^{-i\varphi_{p}}}{\cosh (t/\delta t_{p})},
\label{chirped-pulses-2}
\\
\dot{B}_{p}(t) & =\Delta + \frac{\beta_{p}}{\pi \delta t_{p}} \tanh (t/\delta t_{p}),
\label{chirped-pulses-3}
\end{align}

\noindent
where $\tau=t+z/v_g$,
$\chi_{m,p}(\mathbf{r}_{\perp})=\chi_{0;m,p} f_{g}(\mathbf{r}_{\perp}/a)$. 

The parameters of the control pulses negligibly change  in an optically thin medium, so that we can approximate the Rabi frequency of the control fields:
$\Omega_{m;p}(\mathbf{r},t) = \Omega_{0,m}(\mathbf{r}_{\perp}) a_{p}(t+z/c)$ (where $p=1,2$, $a_{p}(t+z/v_g)=\langle\hat{a}_{p}(z,t)\rangle $ is a classical value).
The maximum CW optical power of the control beam before cryostat is 9 mW.
The parameters $\beta_{p}$ and $\delta t_{p}$  are responsible for the frequency sweep range and the pulse duration, respectively and are chosen in the range $\beta_{p}/\pi=1 \div 7$ and $(2\pi \delta t_{p})^{-1} =  140 \div 400$ kHz. The input pulse, its transmitted portion and the ROSE signal  are 
measured by APD/SPAD1 and APD/SPAD2, respectively, as shown in Fig.\ref{Fig::exp-setup}a).

Fig. \ref{Fig::ROSE+DIAG}a) and Fig. \ref{Fig::ROSE+DIAG}b) show the experimental data on the storage of the weak input light pulse with  the waveguided ROSE protocol. 
Timing of the input signal, two control pulses, and the echo is presented in Fig.\ref{Fig::ROSE+DIAG}a). 
The input signal light pulse with Gaussian waveform and full width half maximum duration of 1 $\mu$s is launched at t=0. 
The rephasing pulses are shown at t = 7.5 $\mu s$ and t = 22.5 $\mu s$ measured by a detector in the reference channel of rephasing beam.
The recovery efficiency 0.5$\%$ of the input pulse  is achieved for a storage time of 30 $\mu s$. 
It should be noted that for the used geometry, the absorption value $\alpha L$ = 0.12 and the coherence time of the optical transition $T_{M}$ = 63 $\mu s$ (x = 1.82),
the maximum efficiency of the ROSE pulse is limited by $\eta_{max}= (\alpha L)^2 \cdot e^{-\alpha L} \cdot e^{-2(4 \tau/T_M)^x}$ with $4\tau$ being the storage time), bearing in mind two ideal control $\pi$ light pulses.
According to this estimation, the maximum achievable efficiency in our optically thin crystal is $\eta_{max}$ = 0.76$\%$.
Below we analyze the experimental data obtained for the ROSE protocol in a waveguide using \eqref{Area_theor_e}, \eqref{parameters_2}, \eqref{pulse_area_s} taking into account the experimental parameters of the medium, signal pulses and modeling the parameters of the control pulses by Eqs.\eqref{control_pulses}-\eqref{chirped-pulses-3}.

\subsection{Theoretical discussion of experimental results}

\begin{figure*}
\centering
\includegraphics[width=\linewidth]{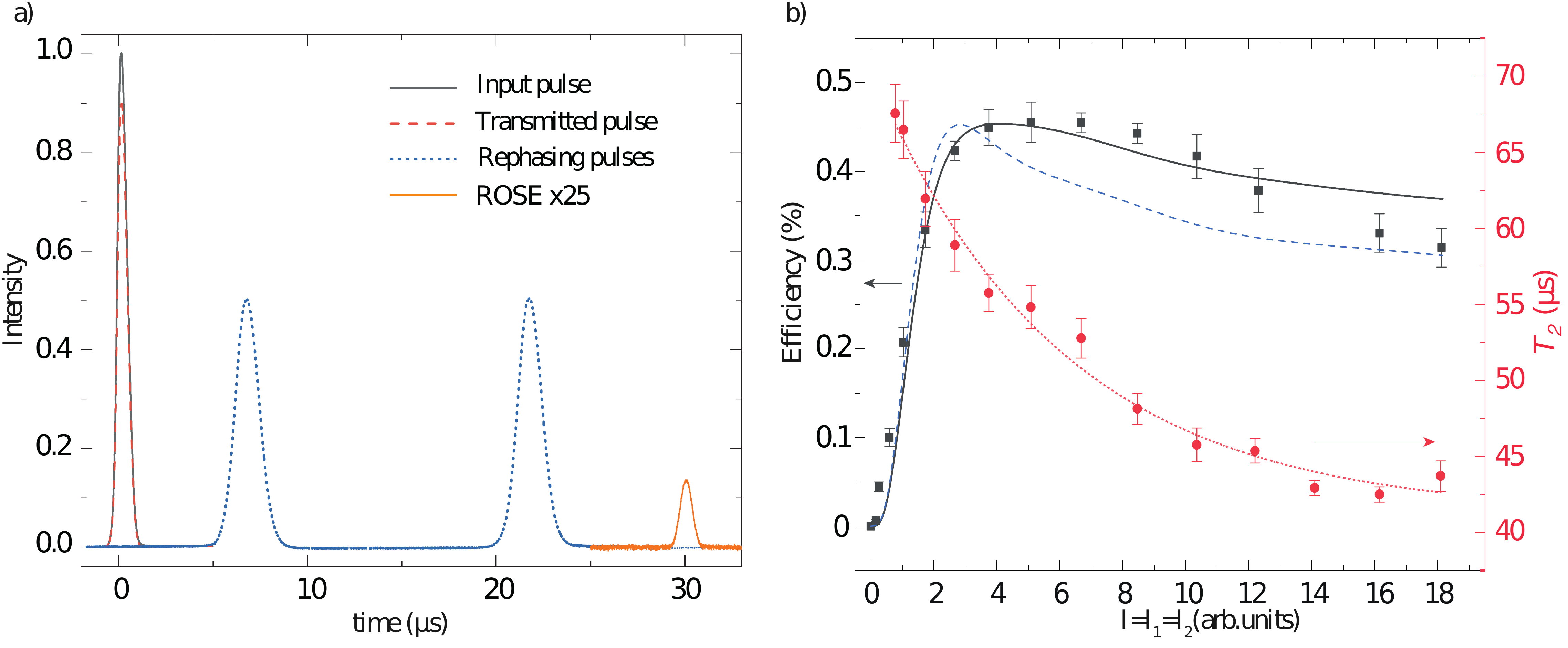}
\caption{\textbf{a)} The signal of the revived of silenced echo magnified by 25 is shown at time 30 $\mu$s (\textcolor{orange}{orange} curve) obtained in the waveguide in Tm$^{3+}$:Y$_3$Al$_5$O$_{12}$ crystal.
The rephasing pulses are shown by the blue dotted curve. 
At t=0 $\mu$s input pulse and transmitted part are shown by black and red dashed curves, respectively. 
\textbf{b)} The retrieval efficiency of input pulse (\textbf{black} squares) and effective coherence time $T_2$ of optical transition (\textcolor{red}{red} circles) of the waveguide optical memory versus intensity of two identical rephasing pulses $I=I_1=I_2$. 
The theoretical fits with Eq. \eqref{solution_f_2} (\textbf{black} solid line) and plane wave theory (\textcolor{blue}{blue} dashed line) are presented.}
\label{Fig::ROSE+DIAG}
\end{figure*}

In our experiment, the inhomogeneous broadening of the resonant optical transition is larger than the spectrum of the control pulses ($\Delta_{in}\gg\delta\omega_{1,2}$, see also Fig. \ref{Figure_1}), and the spectrum width of the signal pulse 
was even smaller ($\delta\omega_s\approx\delta t_s^{-1}<\delta\omega_{1,2}$).
As depicted in the Fig. \ref{Fig::ROSE+DIAG}a), in this regime the temporal shape of the echo signal reproduces the temporal shape of the signal pulse.  
Hence the realized protocol fulfills spectral conditions for the efficient implementation of the photon echo QM \cite{Moiseev2004,Kraus2006,Sangouard_2007}
The reproduction of temporal profile in the echo allows to apply the pulse area approach with
Eqs. \eqref{Area_theor_e}, \eqref{parameters_1}, \eqref{parameters_2} to describe the basic patterns of the realized ROSE protocol.

For a negligibly weak pulse area of the input signal pulse
the probability of exciting the atoms $\mathcal{P}_{11}$ after the first chirped pulse ($p=1$) is calculated using the analytical solution \cite{Hioe1984} (see also Appendix C):
\fla{
&\sin^2\big(\frac{1}{2} \tilde{\Theta}_{m,p}(\mathbf{r}_{\perp})\big)=
\nonumber
\\
&\sin^2\big(\frac{1}{2} \Phi_{m,p}(\mathbf{r}_{\perp})\big)+
\cos^2\big(\frac{1}{2} \Phi_{m,p}(\mathbf{r}_{\perp})\big)
 \tanh^2\big(\frac{\beta}{2}\big),
\label{probability_c}
}
\noindent
where  $\tilde{\Theta}_{m,p}(\mathbf{r}_{\perp})$ is an effective pulse area of chirped pulse for resonant atoms ($\Delta=0$), $p=1,2$,  $\Phi_{m,p}(\mathbf{r}_{\perp})=(\chi_{m,p}^2(\mathbf{r}_{\perp})-\beta^2)^{\frac{1}{2}}$ with the same chirping being used for both control pulses  $\beta_p=\beta$.

Assuming negligible change in the parameters of the control laser pulses in the optically thin medium, Eq. \eqref{Area_theor_e} is adopted for the envelope area of the echo $\theta_{e}$:
\fla{
&\big(\frac{\partial}{\partial z}+\frac{\gamma_w}{2v_g}\big)\theta_{e}=
\sum_{m=1}^{M}\frac{\xi_m}{2}\int_S dxdy \Omega_{0,m} (\mathbf{r}_{\perp} )\cdot
\nonumber
\\
&\big\{2 \Gamma(\mathbf{r},\tau,T_M,\Theta,\beta)
P_m(\mathbf{r}_{\perp},z)+
w_m(\mathbf{r}_{\perp},z)
\Omega_{0,m}(\mathbf{r}_{\perp})\theta_{e}\big\},
\label{Area_theor_weak}
}

\noindent
where $P_m(\mathbf{r}_{\perp},z)$ and $w_m(\mathbf{r}_{\perp},z)$ are given in Eqs. \eqref{parameters_2}, \eqref{pulse_area_s} (see also Appendix C), $\Gamma(\mathbf{r},\tau,T_M,\Theta,\beta)= e^{-(4 \tau/T_M)^x} \exp\{-2\frac{\tau}{T_{s}(\mathbf{r}_{\perp },\Theta,\beta)}\}$ characterizes the phase relaxation of resonant atoms located in a cross-section of the waveguide with a coordinate $\mathbf{r}_{\perp }$;
an additional phase relaxation time $T_{s}(\mathbf{r}_{\perp},\Theta,\beta)$ is determined by the relaxation processes depending on the location of the atom in the waveguide  (pulse area $\Theta$ and $\beta$  determines the degree of atomic excitation in  all atomic groups $m=1,...,M$).

To study the echo pulse area with Eq.\eqref{Area_theor_weak}, it is necessary to independently determine the phase relaxation factor $\Gamma(\mathbf{r},\tau,T_M,\Theta,\beta)$.  
The properties of $\Gamma(\mathbf{r},\tau,T_M,\Theta,\beta)$ are extracted from experimentally measured dependence of the ROSE signal decay at different intensity of the control pulses.
The obtained results shows that the effective phase relaxation time $T_{s}$ of the ROSE signal depends on the intensity of the control pulses, as depicted by the red circles in Fig. \ref{Fig::ROSE+DIAG}b). 
We fit the dependence of effective relaxation time $T_{s}(I,\beta)$  (red dotted line in this figure) on the intensity $I$ of the  control pulses at the given phase modulation $\beta$.
The intensity of the control pulses reduces the phase relaxation time of the optical coherence by about 1.5 times (from 67 to 42 $\mu$s) for the concentration of Tm$^{3+}$  0.01 at.$\%$, which is an order of magnitude less than in the well-known experiments \cite{2014-LP-Thiel}, where the additional attenuation of the echo signal is explained by the influence of the spectral diffusion effect observed earlier.

Taking into account the obtained experimental data, we limit ourselves to the effective constant of phase relaxation $1/T_{s}(I,\beta)$.
Determination of the phase relaxation time $T_{s}(\mathbf{r}_{\perp },\Theta,\beta)$ requires the a more detailed theory that takes into account the features of phase relaxation in an optical waveguide, the inhomogeneous distribution of the intensity of the control fields and the additional experiments, which is beyond the scope of this work and will be carried out somewhere else.
It is also worth noting that the created waveguides remain imperfect and it will be necessary to take into account the influence of the defects created in them.
At the same time, we note that the defects that appear in the waveguide do not dramatically reduce the phase relaxation time \cite{Minnegaliev_2018}, which preserves the hope for further use of the waveguides under study.

If the parameters of the two control laser pulses are identical $\tilde{\Theta}_{m,2}(\mathbf{r}_{\perp},z)=\Tilde{\Theta}_{m,1}(\mathbf{r}_{\perp},z)$, the solution of \eqref{Area_theor_weak} is:

\fla{
\theta_{e}(z,\xi,\tau)=&2\frac{e^{-(4 \tau/T_M)^x}A_{w}(I,\Theta,\xi,\tau)\theta_{s}(0)}{(\alpha_{w}-\alpha)} \cdot
\nonumber
\\
&\big(e^{-\frac{1}{2}\alpha z}-e^{-\frac{1}{2}\alpha_{w}z}\big),
\label{solution_f}
}

\noindent
where $A_{w}(I,\Theta,\xi,\tau)=  \sum_{m=1}^M A_{0,m}(I,\Theta,\xi,\tau)$, $\alpha_{w}=\gamma_{w}+\sum_{m=1}^M\varkappa_{m,w}$ is an absorption coefficient for echo pulse:

\fla{
\varkappa_{m,w}=&\frac {\xi_m}{2}\int_S dxdy\Omega_{0,m}^2 (\mathbf{r}_{\perp} )\cos^2\big( \tilde{\Theta}_{m,1}(\mathbf{r}_{\perp})\big),
\\
A_{0,m}(I,\Theta,\xi,\tau)=& \xi_m 
\exp\{-\frac{2\tau}{T_{s}(I,\beta)}\}
\int_S dxdy\Omega_{0,m}^2 (\mathbf{r}_{\perp} )\cdot
\nonumber
\\ 
&\sin^4\big(\frac{1}{2} \tilde{\Theta}_{m,1}(\mathbf{r}_{\perp})\big).
\label{formules_35}
}
\\
where $A_{0,m}(I,\Theta,\xi,\tau)$ describes the  averaged over the waveguide cross-section response of m-th atomic group, which  contributes to the amplitude of ROSE-signal and $A_{w}(I,\Theta,\xi,\tau)$ gives a total response from all the atomic groups. 
Remarkably, the observed averaged  phase relaxation factor $\exp\{-\frac{2\tau}{T_{s}(I,\beta)}\}$ in Eq. \eqref{formules_35} leads to non-exponential temporal decay.
From Eq. \eqref{solution_f}, we get for the envelope area $\theta_{e}$ at the crystal output ($z=L$) in optically thin atomic ensemble ($\alpha L\ll 1, \alpha_{w}L\ll 1 $):


\fla{
\theta_{e}(L,I,\Theta,\xi,\tau)= e^{-(4 \tau/T_M)^x}\theta_{s}(0) A_{w}(I,\Theta,\xi,\tau)L.
\label{solution_f_2}
}

The function $A_{w}(I,\Theta,\xi,\tau)$ together with $e^{-(4 \tau/T_M)^x}$  describes the influence of the transverse structure of the waveguide light modes, intensity and chirping of control pulses, as well as the intensity dependent phase relaxation $\exp\{-\frac{2\tau}{T_{s}(I,\beta)}\}$.

 The experimental dependence of the ROSE signal on the intensity of two identical control laser pulses is presented in Fig. \ref{Fig::ROSE+DIAG}b).
The experimental data are satisfactorily fitted by $|\theta_{e}(L,I,\Theta,\beta,\tau, \delta t_{p})|^2$ from Eq.\eqref{solution_f_2}  with $\beta/\pi=7$, $\tau=8.5$, $\delta t_{p}=3$ and  Gaussian membrane function  with single type of atomic dipole moment.
The relatively large value of $\beta/\pi=7$ is the best fit to the experimentally realized envelope of the control pulses with their smaller central part of the pulses compared to its fronts.
Similar theory for plane waves, i.e. of the uniform intensity in the cross-section, is presented in Fig. \ref{Fig::ROSE+DIAG}b)
As it is seen from the comparison between the two theoretical curves, the echo signal for the control fields with a plane wave has sharper maximum. The echo signal then decreases faster due to the influence of phase relaxation that is proportional to the intensity of the control field.
The experimentally observed echo signal has a longer plateau in the maximum region, which is consistent with the theoretical curve based on the WPA theorem. 
This behavior of the echo is apparently due to the compensation of growing phase relaxation by an increase in the number of atoms that are excited  by $\pi$ pulses in the cross-section of the waveguide at larger intensities of the control pulses.
The relatively small difference in the theoretical curves is also due to the weak phase relaxation.

We conclude that instantaneous spectral diffusion is mainly responsible for the drop in the efficiency of the echo signal.
There is a slight difference between the experimental data and the theoretical dependance of echo on  intensity of the control pulses.
The difference may be attributed to a number of reasons, including an approximate description of the dependence of phase relaxation on the intensity of the control pulses.
At the same, the quadratic dependence of the echo signal on intensity of the control pulses before the maximum indicates presence of phase modulation in control laser pulses.

\section{Conclusion}

In this work, we derive the waveguide pulse area (WPA) theorem for the interaction of light pulses with an inhomogeneously broadened two-level medium in a single mode optical waveguide.
We present an analytical solution to this theorem for a mode with a Gaussian intensity profile. 
Possibility of $2\pi$ pulses formation for fundamental Gaussian and Gaussian-type (Bessel) light modes was shown.
The formation of these $2\pi$ pulses and their main  properties are compared to prediction of McCall-Hahn area theorem \cite{McCall1969}.
The formation of a $2\pi$ pulses in a single-mode fiber raises the question of the spatial-temporal structure of the resulting  light pulses, which is the topic of special research.

Next, we develop the WPA theorem for studies of photon echo in the single mode waveguide and apply it to analyse the two-pulse (primary) photon echo and the full sequence of generated echo signals after two pulse excitation. 
It is possible to form a sequence of echo signals with a total pulse area of $2\pi$.  
It coincides with the results obtained for the plane waves \cite{HAHN1971, Moiseev2020}.  However, the initial total pulse area of the two exciting pulses should exceed $2\pi$, and not $\pi$ as in case with plane wave.

We use the WPA theorem for the description of the ROSE protocol in a single mode waveguide and, 
for the first time, implement this protocol in an optically thin single-mode laser-written waveguide in Tm$^{3+}$:Y$_3$Al$_5$O$_{12}$ crystal. 
The observed recovery efficiency of input pulse for a storage time of 30 $\mu$s was 0.5 $\%$ that corresponds to  $65.7 \% $ of maximum possible value at given optical depth. 
In these experiments, for the first time, we observed considerable influence of the control laser pulse's intensity on the phase relaxation of ROSE signal in a crystal with the lowest concentration 0.01 \% of $\text{Tm}^{3+}$ ions in a Tm$^{3+}$:Y$_3$Al$_5$O$_{12}$ crystal.
Despite the use of rephasing pulses with amplitude and frequency modulation,  the retrieval efficiency still decreases with an increase in the control pulses intensity.

The theoretical analysis with WPA theorem allows to satisfactorily explain the obtained experimental data.
It was experimentally shown that the waveguide nature of the interaction between light fields and atoms manifests itself in an echo signal even for an optically thin atomic medium.
The decrease in the echo signal with an growth in the intensity of the controlling laser pulses is explained by a decrease in the phase relaxation time, which could be explained by the effect of instantaneous spectral diffusion. 
In order to highly suppress the negative effect of instantaneous spectral diffusion for the implementation of QM on photon echo, one can use the crystals with even lower concentration of the atoms, which is necessary for the QM protocols in high-quality optical resonators \cite{EMoiseev2021}. 
It is worth noting, that not only the presence of pure dephasing and instantaneous spectral diffusion \cite{2014-LP-Thiel,2015-NJP-Dajczgewand} limits the efficiency of the ROSE protocol in the waveguide, but also the difficulties in achieving an ideal control laser $\pi$ pulse providing uniform distribution of working atoms in the waveguide  cross-section.
Fabrication of the waveguides with the active atoms within close proximity of waveguide mode's center is of a great interest for subsequent research.

The performed studies demonstrate the convenience of WPA theorem for describing nonlinear interaction, generation and propagation of $2\pi$  pulses and the photon echo based QM protocols in a single-mode optical waveguides.
Moreover, the developed approach may also be applied for studies of these effects in other waveguides.
It is worth noting that practically all photon echo QM protocols can be implemented in a three-level media, where new opportunities appear for quantum processing with signal fields \cite{Campbell_2014} and for developing new methods for generating quantum states of light \cite{Beavan_2012,Moiseev_2020}.
The derived WPA theorem could be a useful method for accounting for nonlinear effects and related phenomena in various photon echo QM protocols in waveguides with three-level media, where the possibility of quantum storage on long-lived electron-nuclear spin quantum transitions poses new challenges.

\begin{acknowledgments}
 This research was supported by the Ministry of Science and Higher Education of the Russian Federation (Reg. number NIOKTR 121020400113-1).
 S.P.K. and A.A.K. are supported by the Ministry of Science and Higher Education of the Russian Federation on the basis of the FSAEIHE SUSU (NRU) (Agreement No. 075-15-2022-1116).
\end{acknowledgments}

\appendix

\section{Heisenberg-Langevin equations of light and two-level atoms  in single mode waveguide}

For a more general description of the interaction between a light pulse and two-level atomic ensemble in a single waveguide, we will introduce the interaction of  waveguide modes with their environment, where the total Hamiltonian $\hat{H}_0$ in Eq. \eqref{Hamiltonian}  is supplemented by two members and takes the form
$\hat{H}=\hat{H}_0+\hat{H}_{b}+\hat{V}_{bf}$,
where
$\hat{H}_b=\hbar\sum_{n=1}^{N_w}\int d\omega_n\omega_n\hat{b}^\dagger(\omega_n)\hat{b}(\omega_n)$ 
is the  Hamiltonian of the local spatial bath modes ($N_w$ is its number) and 
$\hat{V}_{fb}=\hbar\sum_n f_n\hat{a}^{\dagger}(z_n)e^{-i\beta z_n}\hat{B}_n+h.c.$
(where $\hat{B}_n=\int d\omega_n\hat{b}(\omega_n)$, $f_n$ is a coupling constant of field mode with bath modes at $z_n$, below we assume $f_n=f$ for simplicity) is  interaction of bath modes with the light field modes;
interaction with local reservoir modes leads to the irreversible attenuation of the light modes, in particular, when they are scattered on the inhomogeneities of waveguide walls.

Using  $\hat{H}$, 
we derive a Heisenberg-Langevin equation for the field mode operator $\hat{a}_0(z,t)$ following the derivation of Eq.  \eqref{eq::Asign} and using the well-known input-output formalism \cite{Scully1997,Walls} for calculating the relaxation terms and Langevin forces: 

\fla{
&\left( \frac{\partial}{\partial t} +\frac{\gamma_w}{2}+ v_g \frac{\partial}{\partial z} \right) \hat{a}_0(z,t) =  
\nonumber\\ 
&\frac{i}{2}\sum_{m=1}^{M} \sum^{N_m}_{j=1} \Omega_{0,m} (\mathbf{r}^{j}_{\perp} ) \hat{\sigma}^{j}_{-,m}(t)\delta(z-z_j)+\sqrt{\gamma_w}\hat{b}_{in}(z,t),
\label{eq::Asign-2}
}

\noindent
where Eq. \eqref{eq::Asign-2} differs from Eq.  \eqref{eq::Asign} by an appearance of the relaxation term
$\frac{\gamma_w}{2}\hat{a}_0(z,t)$ and 
Langevin force 
$\hat{b}_{in}(z,t)=-i\sqrt{\frac{2\pi}{\gamma_w}}  f^*\sum_n \hat{B}_n(t\rightarrow -\infty)e^{-i(\omega_n-\omega_0)(t-t_o)}\delta(z-z_n)$
(where $\langle\hat{b}_{in}(z,t)\rangle=0$, $[\hat{b}_{in} (z,t)\hat{b}^{\dagger}_{in}(z',t')]=\delta(z-z')\delta(t-t')$),
coupling constant $\Omega_{0,m}(\mathbf{r}^{j}_{\perp})$ and group velocity $v_g$ are discussed after Eq. \eqref{Hamiltonian}, 
$\gamma_w=2\pi^2 |f|^2 \rho $ is a losses rate of  waveguide modes related to the Langevin force 
 ($\rho=N_w/L$, $L$ - length of the waveguide).

The local operators $\frac{\gamma_w}{2}\hat{a}_0(z,t)$ and $\hat{b}_{in}(z,t)$  describe the light decay in the waveguide and Lagevin forces in the continuous medium, where $\hat{b}_{in}(z,t)$ is determined by the input local fields $\hat{b}(\omega_n,t\rightarrow-\infty)$.
Therefore, by introducing local noise operators $\hat{b}_{out}(z,t)$ at  $t\rightarrow\infty$ similarly to the resonator input-output approach \cite{Walls}, we get the following local relation: 

\fla{
\hat{b}_{in}(z,t)-\hat{b}_{out}(z,t)=\sqrt{\gamma_w}\hat{a}_0(z,t).
\label{input-output}
}
\noindent
The relation \eqref {input-output} is useful for describing the absorption of a light pulse and can be applied to consider echo signals in waveguides where the absorbed signal is localized in the excited (bath) modes of the medium (waveguide) and can be recovered from them by controlled rephasing.
Together with field Eq.\eqref{eq::Asign-2}, we similarly derive Heisenberg-Langevin equations for the resonant atoms \cite{Scully1997}, limiting ourselves only to taking into account the influence of weak phase relaxation: 

\fla{
\frac{\partial \sigma^{j}_{-,m} }{\partial t} = &-(\gamma/2+  i\Delta_{j})  \sigma^{j}_{-,m} 
\nonumber
\\
&-\frac{i}{2} \Omega_{0,m}(\mathbf{r}^{j}_{\perp})\hat{a}_0(z_j,t) \sigma^{j}_{3,m}+\sqrt{\gamma}\hat{F}_{j,m},
\label{eq::Asigma_-2}
\\
\frac{\partial \sigma^{j}_{+,m} }{\partial t} =& -(\gamma/2-i\Delta_{j})  \sigma^{j}_{+,m} 
\nonumber
\\
&+\frac{i}{2}\Omega_{0,m}(\mathbf{r}^{j}_{\perp})\hat{a}_0^{\dagger}(z_j,t)\sigma^{j}_{3,m}+\sqrt{\gamma}\hat{F}_{j,m}^{\dagger},
\label{eq::Asigma_+2}
 \\
\frac{\partial \sigma^{j}_{3,m} }{\partial t} = & - i\Omega_{0,m}(\mathbf{r}^{j}_{\perp})\left[\hat{a}_0^{\dagger}(z_j,t) \sigma^{j}_{-,m} -\hat{a}_0(z_j,t)\sigma^{j}_{+,m}\right].
\label{eq::Asigma_Z2}
}
 The Eqs. \eqref{eq::Asigma_-2}-\eqref{eq::Asigma_Z2} of j-th atom contain only pure dephasing decay constant $\gamma$ with related  Langevin forces $\sqrt{\gamma}\hat{F}_{j,m}(t)$ and $\sqrt{\gamma}\hat{F}_{j,m}^{\dagger}(t)$ (where $\langle\hat{F}_{j}(t)\hat{F}_{j'}^{\dagger}(t')\rangle=\delta_{j,j'}\delta(t-t')$ ) that are assumed to be identical for all the atoms, $\Delta_{j}(t)=\Delta_{j}+\delta\Delta_j(t)$, 
 $\Delta_j$ and $\delta\Delta_j(t)$ are the static and fluctuating frequency offsets of $j$-th atom, where $\Delta_j$ are inhomogeneous broadened by spectral distribution $G(\frac{\Delta}{\Delta_{in}})$,  $\Delta_{in}$ is a linewidth \cite{allen1975}.

Introducing the  density operators of photon number $\hat{I}_p(z,t)=\hat{a}_p^{\dagger}(z,t)\hat{a}_p(z,t)$ and atomic inversion 
$\hat{\sigma}_3(z,t)=\sum_{m=1}^{M} \sum^{N_m}_{j=1} \hat{\tilde{\sigma}}^{j}_{3,m}(t)\delta(z-z_j)$ and  using Eqs.\eqref{eq::Asign-2}, \eqref{eq::Asigma_Z2} we get the following waveguide equation for these operators, describing the behavior of the energy flow:
\fla{
\partial_t(\hat{I}_p+\frac{1}{2}\hat{\sigma}_3)+v_g\partial_z \hat{I}_p =
-\gamma_w \hat{I}_p+ \sqrt{\gamma_w}\hat{M}_{p,b},
\label{photon_number_1}
}
\noindent
 where the two-particle operator  $\hat{M}_{p,b}=\hat{a}_p^{\dagger}(z,t)\hat{b}_{in}(z,t)+h.c.$,
 the  quantum mechanical average of which is zero due to the independent nature of the noise operator $\hat{b}_{in}(z,t)$ and $\langle\hat{b}_{in}(z,t)\rangle=0$.
The Eq.\eqref{photon_number_1} is valid for arbitrary constants of light-atom interaction, so it hiddens a specific dynamics under studied.
In the case of spatially localized $2\pi$-excitation  ($2\pi$-echo sequence in Section III), the analytical solution of the field and atomic operators takes the form 
$\hat{I}_p(z,t)=\hat{I}_p(\xi)$, 
$\hat{\sigma}_3(z,t)=\hat{\sigma}_3(\xi)$,
which satisfy the equation:
\fla{
\hat{\sigma}_3(\xi)=\hat{\sigma}_3(-\infty)+
2(\frac{v_g}{\tilde{v}_g}-1)
(\hat{I}_p(\xi)-\hat{I}_p(-\infty)),
\label{photon_number_2}
}
\\
under the conditions of negligible relaxation, where  $\hat{\sigma}_3(-\infty)$ and $\hat{I}_p(-\infty)$ are the initial atomic and field operators, $\xi=t-\frac{z}{\tilde{v}_g}$, $\tilde{v}_g$ is a group velocity of the light-atoms (polariton) packet which is  less than initial group velocity $v_g$ of light without atomic medium and  should be found as a result of solving a system of Eqs.\eqref{eq::Asign}-\eqref{eq::Asigma_Z} (i.e. Eqs. \eqref{eq::Asign-2}-\eqref{eq::Asigma_Z2} not containing relaxation terms).
It is worth noting the exact analytical solution of the quantum Maxwell-Bloch   equations \eqref{eq::Asign}-\eqref{eq::Asigma_Z} with the identical constants of the light-atom interaction \cite{Rupasov1982}, demonstrating the possibility of quantum $2\pi$ solitons.
However, the Eqs. \eqref{photon_number_1}, \eqref{photon_number_2} also indicate the possibility of propagation of  light pulses at a constant group velocity even in the presence of a spread of the constants of light-atom interaction.

Eqs. \eqref{eq::Asign-2}-\eqref{eq::Asigma_Z2} should be solved for the stages of storage and retrieval of the fields in arbitrary quantum states. 
In the main part of work we focus on the consideration of the classical light fields, moving from the field operators $\hat{a}_m(z,t)$  to the c-numbers 
$a_{p}(z,t)=e^{-i\phi}\langle\hat{a}_{p}(z,t)\rangle=e^{i\phi}\langle\hat{a}_{p}^{\dagger}(z,t)\rangle$, $p$ $(s;1;2;e)$ index indicating signal, controls and echo pulses, see comments to Eq. \eqref{eq::Asigma_Z}
and omitting the Langevin forces ($\langle\hat{b}_{in}(z,t)\rangle=0$).
At the same time, it should be noted that the semi-classical Maxwell-Bloch equations can be applied to describe the coherent interaction between light fields weakened to a single-photon level with atomic media (see also comments after Eq.\eqref{eq::Asigma_Z}). 
A number of basic properties of optical quantum memory protocols, such as the efficiency of signal pulse storage the accuracy of its retrieval, can be also analyzed.

\section{ROSE protocol in optically dense waveguide}

Here we assume that the signal and control fields have the same carrier frequency and do not have additional phase modulation. 
Using $w_m^{re}(\mathbf{r}_{\perp},z)$ and $P_m^{re}(\mathbf{r}_{\perp},z)$ in \eqref{Area_theor_e}  together with phase relaxation function $\Gamma(\mathbf{r},\tau,T_M,\Theta_{1,2})$ we can perform general analysis of nonlinear pattern in ROSE-pulse area behavior.
In the simplest case of negligible spatial dependence of phase relaxation (i.e., $\Gamma(\mathbf{r},\tau,T_2,\Theta_{1,2})= \Gamma(\tau,T_2,...)$), we can perform integration and subsequent analytical  calculation of \eqref{Area_theor_e}  similar to \eqref{Envelope_area_с2} and to get the following equation for the envelope area of ROSE pulse:

\fla{
\big(\frac{\partial}{\partial z}+\frac{\gamma_w}{2v_g}\big)\theta_{e}=
&\sum_{m=1}^{M}\frac{\varkappa_m}{\Omega_{m}}
\{\frac{\Gamma(\tau,T_M,...)}{4} I_{s,m}(\theta_{s},\theta_{1}, \theta_{2},\theta_{e})
\nonumber\\
&-S_m(\theta_{e};\theta_{1};\theta_{2})
\},
\label{Envelope_echo-area}
}

\noindent
where a source of echo signal:

\fla{
& I_{s,m}(\theta_{s},\theta_{1}, \theta_{2},\theta_{e})=
\nonumber
\\
&\frac{1}{2}\Omega_{m}\theta_{s}(z)+S_m(\theta_{s};\theta_{e})+
S_m(\theta_{s};\theta_{1};\theta_{2})
\nonumber\\
&-S_m(\theta_{s};\theta_{e};\theta_{1})-
S_m(\theta_{s};\theta_{e};\theta_{2})+S_m(\theta_{s};\theta_{e};\theta_{1};\theta_{2}),
\label{I_s,1,2,e}
}

\noindent
envelope areas $\theta_{s,1,2,e}(z)$ are all the functions of $z$, we also taken into account that the signal pulse is weak   $\Omega_{m}\theta_{s}(z)\ll\pi$,

\fla{
&S_m(\theta_{1};\theta_2;\theta_3)=
\nonumber
\\
&\frac{1}{4}\sum_{n=1}^{2}\sum_{m=1}^{2}\frac{\sin^2{\big(\Omega_{m}(\theta_{1}+(-1)^n\theta_{2}+(-1)^m\theta_{3})/2\big)}}
{\Omega_{m}(\theta_{1}+(-1)^n\theta_{2}+(-1)^m\theta_{3})/2},
\nonumber
\\
&S_m(\theta_{1};\theta_2;\theta_3;\theta_4)=\frac{1}{8}\sum_{n=1}^{2}\sum_{m=1}^{2}\sum_{p=1}^{2}
\nonumber
\\
&\frac{\sin^2{\big(\Omega_{m}(\theta_{1}+(-1)^n\theta_{2}+(-1)^m\theta_{3}+(-1)^p\theta_4)/2\big)}}
{\Omega_{m}(\theta_{1}+(-1)^n\theta_{2}+(-1)^m\theta_{3}+(-1)^p\theta_4)/2}.
\label{S_1,2,3}
}
\\
Eq. \eqref{Envelope_echo-area} is still a nonlinear equation on $\theta_e$. 
In the case of atomic medium with limited optical density and weak input signal pulse, this equation reduces to the linear equation where $\Omega_{m}\theta_{e}(z)\ll\pi$.
Neglecting the terms with $(\Omega_{m}\theta_e)^2\ll \pi$ in \eqref{Envelope_echo-area}, we get:

\fla{
\big(\frac{\partial}{\partial z}+\frac{\gamma_w}{2v_g}\big)\theta_{e}=&
\sum_{m=1}^{M}
\{\frac{\varkappa_m}{4\Omega_{m}}
\Gamma(\tau,T_2,...) I_{s,m}(\theta_{s},\theta_{c1},\theta_{c2})
\nonumber\\
&-\frac{1}{2}
\varkappa_m(\theta_{c1};\theta_{c2})\theta_{e}
\},
\label{Envelope_area_ROSE-echo-w}
}

\noindent
where $I_{s,m}(\theta_{s},\theta_{1},\theta_{2})\equiv I_{s,m}(\theta_{s},\theta_{1},\theta_{2},0)$: 

\fla{
I_{s,m}(\theta_{s},\theta_{1},\theta_{2})=&
\Omega_{m}\theta_{s}-S_m(\theta_{s};\theta_{1})-
\nonumber\\
& S_m(\theta_{s};\theta_{2})+
2S_m(\theta_{s};\theta_{1};\theta_{2}),
\label{formula_3}
}

\noindent
and absorption coefficient for echo signal due to the interaction with $m$-th atomic group is:
\fla{
&\varkappa_m(\theta_{1};\theta_{2})=\varkappa_m\sum_{n=1}^{2}
\big\{\frac{\sin{\big(\Omega_{m}(\theta_{1}+(-1)^n\theta_{2})\big)}}{\Omega_{m}(\theta_{1}+(-1)^n\theta_{2})}
\nonumber\\
&-\frac{\sin^2{\big(\Omega_{m}(\theta_{1}+(-1)^n\theta_{2})/2\big)}}{2(\Omega_{m}(\theta_{1}+(-1)^n\theta_{2})/2)^2}
\big\}.
\label{formula_4_dop}
}

Eq. \eqref{Envelope_area_ROSE-echo-w} is used in the main text.

\section{Excitation of atoms by controlling chirped pulses}
 
Here following the work \cite{Hioe1984}, we consider the excitation of two-level atoms by two chirped control pulses.
In the interaction picture, the equations for atomic  amplitudes  in the ground ${C}_2(t)$ and exited ${C}_1(t)$ states are: 
\begin{equation}
	\begin{pmatrix}
		\dot{C}_1(t) \\
        \dot{C}_2 (t)      
	\end{pmatrix} =\frac{i}{2}
    \begin{pmatrix}
    	0 & \Omega(t) \\
       \Omega^*(t)  & 0
    \end{pmatrix}
    \begin{pmatrix}
		C_1(t) \\
        C_2 (t)      
	\end{pmatrix},
    \label{eq:CMatrixSet}
\end{equation}
with initial state  ${C}_2(t\rightarrow - \infty)=1$,  ${C}_1(t\rightarrow - \infty)=0$, where

\begin{align}
\Omega(t)&=\Omega_1(t-t_1)+\Omega_2(t-t_2)
\nonumber
\\
&=(\dot{A}_1 (t) e^{-i B_1(t)}+\dot{A}_2(t) e^{-i B_2(t)}) e^{-i \Delta (t-t_0)},
\label{Omega_1_2}
\end{align}

\begin{align}
\dot{A}_{p}(t) & = \frac{\chi_{p}}{\pi \tau_{p}} \frac{e^{-i\varphi_{p}}}{\cosh [(t-t_{p})/\tau_{p}]}, 
\label{A_p}
\\
\dot{B}_{p}(t) & = \frac{\beta_{p}}{\pi \tau_{p}} \tanh [(t-t_{p})/\tau_{p}].
\label{B_p}
\end{align}
where index $p=1,2$ corresponds to the 
1-st, 2-nd control pulse, ${A}_{p}(t)$  is a Rabi frequency of the chirped pulses. 
We assume that the Rabi frequency is also a function of transverse coordinate $\mathbf{r}_{\perp}$:  $\chi_p(\mathbf{r}_{\perp})=\chi_{p,0}f (\mathbf{r}_{\perp})$; $\varphi_{p}$ is constant phases;   ${B}_{p}(p)$ describes frequency modulation; $\Delta$ is a atomic frequency offset, $\tau_{p}$ is a temporal duration of $p$ control pulse, $\chi_{p}$ and $\beta_{p}$ are the amplitude and frequency chirp of $p$-th control pulse, $f (\mathbf{r}_{\perp})$ is a spatial shape in the cross-section of light beam.
Below, we have taken into account a Gaussian membrane function of the Rabi frequency for each control field: 
$\chi_p(\mathbf{r}_{\perp})=\chi_{p,0}f_{g}(\mathbf{r}_{\perp})$.

By taking into account that the chirped pulses are not overlapped with each other ($\tau_{1,2}\ll t_1-t_0,t_2-t_1$), we can solve the Eqs. \eqref{eq:CMatrixSet} one by one. 
For each pulse the solution  of \eqref{eq:CMatrixSet} can be written in terms of Hypergeometric functions where
it is convenient writing  $C_p(t)$ via the new variable:

\fla{
z_1(t) = \frac{1}{2}(1+\tanh[(t-t_1)/\tau_1 ]),
}
i.e. $C_p(t)= \tilde{C}_p(z_1)$ and 
determining the phase of the first light pulse $\varphi_1$ by setting $B_1(t)=\int_{t_0}^t dt'\dot{B}_{1}(t')$.
The atomic amplitudes after the action of first control pulse (starting  at $t=t_0$) is described by
\begin{align}
	\tilde{C}_1 (z_1) & =  A_1^{(12)} z_1^{1-c_1} \mathcal{F}_1^{(1)}(z_1), \\
     \tilde{C}_2 (z_1) & = C_2 (t_0) \mathcal{F}_1^{(2)*}(z_1),
    \label{Solut:C_1,2}
\end{align}
where we introduced an abbreviated notation for the hypergeometric functions:
\begin{align}
	\mathcal{F}_p^{(1)}(z_p)&=F(a_p+1-c_p,b_p+1-c_p,2-c_p,z_p), 
	\\
\mathcal{F}_p^{(2)*}(z_p)&=F^*(a_p,b_p,c_p,z_p)=F(a_p^*,b_p^*,c_p^*,z_p),
    \label{F-functions}
\end{align}
\begin{equation}
 	A_1^{(12)}= \dfrac{i\chi_1 e^{-i\varphi_{1}}}{2\pi (1-c_1)} C_2(t_0) \exp\{i(\Delta-\frac{\beta_1}{\pi \tau_1})(t_0-t_1)\},
    \label{eq:a2Initial}
\end{equation}

Values $a_p,b_p,c_p$ can be found in \cite{Hioe1984}:

\begin{align}
 	 a_p & = \dfrac{1}{2 \pi } \left[(\chi_p^2-\beta_p^2)^{\frac{1}{2}} - i\beta_p \right], \\
    b_p & = \dfrac{1}{2 \pi } \left[- (\chi_p^2-\beta_p^2)^{\frac{1}{2}} - i\beta_p \right], \\
    c_p & = \dfrac{1}{2}\left[ 1+ i\dfrac{\pi\Delta\tau_p - \beta_p}{\pi}\right].
\end{align}
Interaction with the first pulse ends for $t-t_1\gg\tau_1$ when $z_1\rightarrow 1$ and the amplitudes $\tilde{C}_{1,2} (z=1)$ are set unchanged.   
Probability to find atom in the excited state after the action of the first pulse (for the initial state $|C_2(t_0)|\cong1$ in \eqref{Solut:C_1,2}) is:

\begin{equation}
\mathcal{P}_{11}(t_1)\equiv P_{11} (t\gg t_1+\tau_1) =|\tilde{C}_1(1)|^2=\sin^2(\Theta_1/2),    
\label{Theta}
\end{equation}

\noindent
where 

\begin{align}
&\dfrac{\chi_p^2 |\mathcal{F}_p^{(1)}(1)|^2}{4\pi^2 |1-c_p|^2} =\sin^2(\Theta_p/2)=
\nonumber
\\
&\frac{\sin^2(\Phi_p/2)Ch^2(\beta_p/2)+\cos^2(\Phi_p/2)Sh^2(\beta_p/2)}{Ch(\frac{\beta_p+\pi\Delta\tau_p}{2})Ch(\frac{\beta_p-\pi\Delta\tau_p}{2})},
\label{Probability_p}
\end{align}
where  $\Phi_p=(\chi_p^2-\beta_p^2)^{\frac{1}{2}}$  and  \eqref{Theta}, \eqref{Probability_p} are also used to define the nutation angle of two-level atom $\Theta_{p=1}=\Theta_1(\Delta,\beta_1,\Phi_1)$  caused by the action of  the first pulse.
By using \eqref{Probability_p}, we can easily calculate $|\mathcal{F}_1^{(2)}(1)|^2$ and  the probability of atom to stay on the ground level  $P_{2} (t\gg t_1+\tau_1)=|C_2(t\gg t_1)|^2
=|\tilde{C}_2(1)|^2=|\mathcal{F}_1^{(2)}(1)|^2=1-|\tilde{C}_2(1)|^2=\cos^2(\Theta_1/2)$, so we get

\fla{
C_2(t\gg t_1)=&\mathcal{F}_1^{(2)*}(1)=
\nonumber\\
&\cos(\Theta_1/2)\exp\{-i\zeta_1(\Delta,\chi_1,\beta_1,\tau_1)\},
\label{phase_chi}
}

\noindent
where $\zeta_1(\Delta,\chi_1,\beta_1,\tau_1)=-i\ln\{\mathcal{F}_1^{(2)}(1)/\cos(\Theta_1/2)\}$.

Interaction with the second control pulse occurs at $t_2$  with sufficiently large time delay after the first control pulse $t_2-t_1\gg\tau_1$.
Let's chose a moment of time in the middle between the control pulses $t_m=\frac{t_1+t_2}{2}$.
Using \eqref{eq:CMatrixSet} for $t>t_m$ and moving to new amplitudes $C_{2,0}(t)=C_2(t)\exp\{-\frac{i}{2}\Delta(t_m-t_0)\}$, $C_{1,0}(t)=C_1(t)\exp\{\frac{i}{2}\Delta(t_m-t_0)\}$, 
we can describe the interaction with the second pulse similarly to the interaction with the first pulse. 
Here we get the following system of equations for atomic amplitudes $C_{1,0}(t)$ and $C_{2,0}(t)$ ($t>t_m$), 
which takes into account the phase change of the second pulse at the moment of the beginning of interaction with the atom:

\begin{equation}
	\begin{pmatrix}
		\dot{C}_{1,0}(t) \\
        \dot{C}_{2,0}(t)      
	\end{pmatrix} =\frac{i}{2}
    \begin{pmatrix}
    	0 & \Omega_2(t-t_m) \\
       \Omega_2^*(t-t_m)  & 0
    \end{pmatrix}
    \begin{pmatrix}
		C_{1,0}(t) \\
        C_{2,0} (t)      
	\end{pmatrix},
    \label{eq:CMatrixSet-2}
\end{equation}
with initial amplitudes:
\fla{
{C}_{2,0}(t_m)=\tilde{C}_{2}(1)\exp\{-\frac{i}{2}\Delta(t_m-t_0)\},
\\ 
{C}_{1,0}(t_m)=\tilde{C}_{1}(1)\exp\{\frac{i}{2}\Delta(t_m-t_0)\},
}

\noindent
where $\Omega_2(t-t_m)$ is given in Eqs. \eqref{Omega_1_2}-\eqref{B_p},
and the constant phase of the second light pulse $\varphi_2$. 

Similarly to the interaction with the first pulse, we move to the new time variable scale $z_2 = \frac{1}{2}(1+\tanh[(t-t_2)/\tau_2 ])$ and find solutions of ${C}_{p,0}(t>t_m)=\tilde{C}_{p,0}(z_2)$ for the initial conditions $\tilde{C}_{p,0}(0)={C}_{p,0}(t_m)$: 
 
\begin{align}
\tilde{C}_{1,0}(z_2) = &{C}_{1,0}(t_m) 
\mathcal{F}_2^{(2)}(z_2)
+A_2^{(12)} z_2^{1-c_2}
\mathcal{F}_2^{(1)}(z_2),
 \\
\tilde{C}_{2,0}(z_2) =  &{C}_{2,0}(t_m) 
{F}_2^{(2*)}(z_2)
+ A_2^{(21)} z_2^{1-c_2^*}
{F}_2^{(1*)}(z_2),
    \label{solution:ampl}
\end{align}
where

\begin{align}
A_2^{(12)}&= \dfrac{i\chi_2 e^{-i\varphi_{2}}}{2\pi (1-c_2)} C_{2,0}(t_m) \exp\{i(\Delta-\frac{\beta_2}{\pi \tau_2})(t_m-t_2)\},
  \\
A_2^{(21)}&= \dfrac{i\chi_2 e^{i\varphi_{2}}}{2\pi (1-c_2^*)} C_{1,0}(t_m) \exp\{-i(\Delta-\frac{\beta_2}{\pi \tau_2})(t_m-t_2)\}.
\label{s-matrix}
\end{align}

The probability  of the atomic excitation $\mathcal{P}_{11}(t_2)\equiv P_{11}(t\gg (t_2+\tau_2))=|\tilde{C}_{1,0}(z_2=1)|^2$ after the interaction with two control pulses will be:  

\begin{align}
\mathcal{P}_{11}(t_2)=&  |\tilde{C}_{1,0}(1)|^2  = |{C}_{1,0}(t_m) \mathcal{F}_2^{(2)}(1)|^2+
\nonumber
\\
+&
|A_2^{(12)} \mathcal{F}_2^{(1)}(1)|^2+
\left(\delta \mathcal{P}_{coh}(t_2) +C.C.\right),
\end{align}

\noindent
where we have:

\begin{align}
&|{C}_{1,0}(t_m) \mathcal{F}_2^{(2)}(1)|^2 = 
\sin^2(\Theta_1/2)\cos^2(\Theta_2/2),
\\
&|A_2^{(12)}\mathcal{F}_2^{(1)}(1)|^2=
\sin^2(\Theta_2/2)\cos^2(\Theta_1/2),
\end{align}
\\
and interference term:

\begin{align}
&\delta \mathcal{P}_{coh}(t_2) = {C}_{1,0}(t_m) A_2^{(12)*}\mathcal{F}_2^{(2)}(1)=
\nonumber \\
&\exp\{i[\delta\varphi+\tilde{\varphi}_{2,1}+\Delta(t_2-t_1)]\}\cdot
\nonumber
\\
&\frac{\chi_1\chi_2 
\mathcal{F}_1^{(1)}(1)
\mathcal{F}_2^{(1)*}(1)
}{4\pi^2(1-c_1)(1-c_2^*)}
\mathcal{F}_1^{(2)}(1)
\mathcal{F}_2^{(2)}(1), 
\label{interf_term}
\end{align}

\noindent
where $\delta\varphi$ is a random phase that is acquired by an atom during the time interval $t_2-t_1$ of free evolution due to the interaction with bath fields,
$\tilde{\varphi}_{2,1}=\tilde{\varphi}_2-\tilde{\varphi}_1$, $\tilde{\varphi}_p$ - constant phases of p-th control pulses are defined by setting $B_{1,2}(t)=\int_{t_0,t_m}^t dt'\dot{B}_{2}(t')$ as follows: $\tilde{\varphi}_1=\varphi_1+\frac{\beta_1}{\pi\tau_1}(t_0-t_1)$, $\tilde{\varphi}_2=\varphi_2+\frac{\beta_2}{\pi\tau_2}(t_m-t_2)$.
In \eqref{interf_term} we also used:
${C}_{1,0}(t_m)=\tilde{C}_{1}(1)\exp\{\frac{i}{4}\Delta(t_m-t_0)\}$, $\tilde{C}_1 (1) =  A_1^{(12)} \mathcal{F}_1^{(1)}(1)$, and ${C}_{2,0}(t_m)=\tilde{C}_{2}(1)\exp\{-\frac{i}{4}\Delta(t_m-t_0)\}$, $\tilde{C}_2 (1)= \mathcal{F}_1^{(2)*}(1)$.

In the particular case of identical two control pulses ($\chi_1=\chi_2, \beta_1=\beta_2,\tau_1=\tau_2$ and $\Theta_1=\Theta_2$, respectively), we get

\fla{
&\mathcal{P}_{11}(t_2;\Theta_2=\Theta_1) = 
\nonumber \\
&\frac{1}{2} \sin^2\Theta_1\left\{1+\cos\left[\delta\varphi+2\zeta_1+\Delta(t_2-t_1)\right]\right\}, 
\label{Theta_2}
}
where $\zeta_1$ is given in Eq. \eqref{phase_chi} and 
 the durations of the control pulses are defined by $\tau_{1,2}=\delta t_{1,2}$ in the main text.

\bibliography{apssamp} 

\end{document}